\documentclass[runningheads,fleqn,epsfig]{svmult}

\usepackage{makeidx}   
\usepackage{graphicx}  
\usepackage{subeqnar}  
\usepackage{multicol}  
\usepackage{physmult}  
\makeindex             

\newcommand{\be}{\begin{eqnarray}}
\newcommand{\ee}{\end{eqnarray}}

\begin{document}
\title*{Nonperturbative Phenomena\\ and Phases of QCD }
\author{  Edward V. Shuryak}
\authorrunning{ Edward V. Shuryak}
\maketitle              

\section{ Introduction}
\subsection{An outline}

The lectures provide brief
 overview of what  we have learned about QCD vacuum,
 hadrons and hadronic matter
during the last 2 decades. 
Systematic description of  the topics cover would need
a large book, not lecture notes.
Some material is available in reviews written in more technical way,
for instantons and chiral symmetry breaking those are
\cite{SS_98,Diakonov},
for correlators, OPE etc \cite{Shu_93}. A lot of other material is
recent: one can only consult the original papers.

 In these lectures there are not so many formulae.
 I tried to clarify the main
physics point, the main questions
debated today, and show few recent examples.
Admittedly, the
 title of these lectures is very general: but
they cover a lot of different phenomena. 
 We will  start with the QCD {\em vacuum structure},
move to{\em hadronic structure}, discuss {\em phases} of hot/dense QCD and eventually
consider {\em high energy collisions} of hadrons and heavy ions.

The main line in all discussion would be a
systematic use of semiclassical methods, specifically the instantons. The reasons for that are: (i) They are the only truly
non-perturbative effects understood by now; (ii) They lead to large
and probably even dominant effects in many cases; (iii) Due to
progress during the last decade, we have near-quantitative theory
of instanton effects, solved numerically to {\em all orders} in the so 
called 't Hooft interaction.

Although we still do not understand confinement, its companion problem -
 chiral symmetry breaking in the QCD vacuum - is now understood to a
significant degree. Not only we have simple qualitative understanding
of where those quasi-zero modes of the Dirac operator come from, but
we can calculate their density, space-time shape and eventually
 {\bf QCD correlation functions} with surprising accuracy.
So, in a way, the problem of
 {hadronic structure} is nearly solved for light-quark hadrons\footnote{
Medium-heavy-quark ones, such as $\bar c c,\bar b b$ do care about
confining potential, while very-heavy quarkonia need only the Coulomb
forces.}.
 
  As we will see below, 
 the QCD phase diagram can also be well understood 
in the instanton framework. The boundaries of
three basic phases of QCD: (i) hadronic
phase,
(ii) Quark-Gluon Plasma (QGP) and (iii) Color Superconductor (CS)
phases
appear as a balance between three basic pairing channels, being
(i) attraction in scalar colorless
$\bar q q$ channel; (ii) instanton-antiinstanton
pairing induced by light quark exchanges; and (iii)  attraction in
scalar but
colored $q q$ channels.

The last part deals with heavy ion collisions: those are related to
the rest of the lectures since this is how we try to access hot QCD
experimentally. We will discuss first results coming from RHIC,
show that matter produced seems to behave macroscopically
(namely, hydrodynamically) with proper Equation of State. We will also
try to connect rapid onset of QGP equilibration  with
existing perturbative and non-perturbative estimates.

\subsection{Scales of QCD}
  Let me start with an introductory discussion of various ``scales"
of  non-perturbative QCD. The major reason I do this is the
following: some naive simplistic ideas we had in the early
days of QCD, in the 70's,  are still alive today.  I would
 strongly argue against
the picture of  non-perturbative objects as some structure-less
fields with typical momenta of the order of $p\sim \Lambda_{QCD}\sim (1 \, fm)^{-1}$. In the  mid-70's people considered
hadrons to be  structure-less 
``bags" filled with near-massless perturbative quarks, with  
mild non-perturbative effects appearing 
at its boundaries and confining them at the scale of 1 \, fm.

One logical consequence of this picture
would be applicability of 
 the derivative expansion of the non-perturbative fields
or Operator Product Expansion (OPE), the basis of
the  QCD sum rules. 
  However, after the first successful applications of the method
\cite{SVZ} rather
serious problems\cite{NSVZ} have surfaced.  
All spin-zero channels
(as we will see, those are the ones directly coupled to instantons)
related with quark or gluon-based operators alike,
 indicate unexpectedly large non-perturbative effects and
deviate from the OPE predictions at very small distances.
 
  It provided a very important lesson:  
{\em the non-perturbative fields  form structures
with sizes significantly smaller than 1 \, fm} and local
field strength much larger than $\Lambda^2$. Instantons are one of them:
 in order to
describe many of  these phenomena in a consistent way one needs 
 instantons of small size \cite{Shu_82} $\rho \sim 1/3 \, fm$.
We have direct confirmation of it from the lattice, but not real
understanding of why there are no large-size instantons.

Furthermore, the instanton  is not the only such small-scale gluonic object.
We also learned from the lattice-based works that QCD flux tubes
(or confining strings) also have small radius,
only about $r_{string}\approx 1/5  \, fm$.
So, all hadrons (and clearly the QCD vacuum itself) have a
 $substructure$, with ``constituent quarks'' generated by instantons
connected by such flux tubes.

Clearly this substructure should play an important role 
in hadronic physics. We would like to know why the usual quark
model has been so successful in spectroscopy, and why so little
of exotic states have been seen. Also,  high
energy hadronic
collisions must tell us a lot about substructure, since the famous
Pomeron  also belongs to a list of those surprisingly small
non-perturbative
objects.

At the opposite end of the spectrum, people have found that QCD seem
to have also surprisingly {\em small energy/momentum scale}, several
times lower
than $\Lambda$. It was found that behavior of the so called
``quenched'' and true QCD is very different, but only if
the quark mass is below some scale of the order of 20-50 MeV.
As we will see below, this surprising low scale has been explained by
properties of the instanton ensemble. 

\section{  Chiral symmetry breaking and instantons}
\subsection{Brief history}

Let me start around 1961, when the ideas about chiral symmetry and
what it may take to break it spontaneously have appeared. 
The NJL model \cite{NJL} was the first microscopic
model which attempted  to derive  dynamically
the properties of chiral symmetry breaking  and 
pions, starting from some {\em hypothetical
 4-fermion  interaction}. 
\be L_{NJL}=G(\vec\pi^2+\sigma^2) \ee
where $\pi,\sigma$ denote the corresponding scalar isovector and
scalar isoscalar currents. 

Let me make few  comments about it.\\
(i) It was the first bridge between the BCS theory of superconductivity and 
quantum field theory, leading the way to the Standard Model.
 It first showed that the vacuum can be
 truly nontrivial, a superconductor of a kind, with the mass
 gap $\Delta$=330-400 MeV, known as ``constituent
quark mass".\\
 (ii) The NJL model has  2 parameters: the strength
of its 4-fermion interaction G and the cutoff $\Lambda\sim .8-1 GeV$.
The latter regulates the loops (the model is non-renormalizable,
which is OK for an effective theory) and is directly the ``chiral scale''
we are discussing. We will relate $\Lambda$ to the typical instanton size
$\rho$,
and G to a combination $n\rho^2$ of the size and density of instantons.
\\
(iii)   One non-trivial prediction of the NJL
model was a 
the mass of the scalar is $m_\sigma\approx 2 m_{const.quark}$.
 Because this state
is the P-wave in non-relativistic language, it means that there is
strong attraction which is able to compensate
exactly for rotational kinetic energy. For decades simpler hadronic models
failed to get this effect, and even now
 spectroscopists still argue that this (40-year-old!)
result is incorrect. However, lattice results 
in fact show that it is exactly
right and theoretically understood by instantons.
 Moreover, the phenomenological sigma meson
is being revived now, so possibly it will even get back to its proper place
in Particle Data Table, after  decades of absence.

Let me now jump to instantons. We will show below that they generate
quite specific 4-fermion 't Hooft interaction \cite{tHooft} 
(for 2-flavor theory: for
pedagogical reasons we ignore strange quarks altogether now).
Furthermore, its Lagrangian includes the NJL one, but it also has
2 new terms:
\be L_{tHooft}=G(\vec\pi^2+\sigma^2-\eta^2-\vec\delta^2) \ee
with isoscalar pseudoscalar $\eta$ and isovector scalar $\vec\delta$.
T'Hooft's minus sign is crucial here: it shows that the axial U(1) symmetry
(e.g. rotation
of sigma into eta) is $not$ a symmetry. That is why $\eta$
(actually$\eta'$ if strangeness is included)   is $not$ 
massless Goldstone particle like a pion. 

 The most important next development happened in 1980's: it has been
shown
in \cite{Shu_82,DP_86} that instanton-induced interaction does break
$spontaneously$ the $SU(N_f)$ chiral symmetry.  
 Unlike the NJL model,  the instanton-induced interaction 
has a natural cut-off parameter $\rho$, and the coupling constants 
are not free parameters, but determined by a physical quantity, 
the instanton density. That eventually allowed to solve in all
orders in 't Hooft interaction, and get quantitative
results, see \cite{SS_98}.

\subsection{General things about the instantons}

I would omit from this paper  general things about the instantons,
well covered elsewhere. Let me just briefly mention that
the topologically-nontrivial 4d solution was found by 
Polyakov and collaborators in\cite{BPST},
and soon it was interpreted as semi-classical tunneling
between topologically non-equivalent vacua. The name itself
was suggested by t Hooft, meaning ``existing for an instant''.
  Formally, instantons appear in the context of the semi-classical 
approximation to the (Euclidean) QCD partition function 
\be
\label{Z_QCD}
 Z  =  \int DA_\mu \,\exp(-S)\, \prod_f^{N_f} \det 
  \left( D\!\!\!\!/\,+m_f \right), \\
      S\,=\, \frac{1}{4g^2} \int d^4x\, G^a_{\mu\nu} G^a_{\mu\nu}.
\ee
Here, $S$ is the gauge field action and the determinant of the 
Dirac operator $D\!\!\!\!/\,= \gamma_\mu(\partial_\mu-iA_\mu)$
accounts for the contribution of fermions. In the semi-classical 
approximation, we look for saddle points of the functional integral 
(\ref{Z_QCD}), i.e. configurations that minimize the classical 
action $S$. This means that saddle point configurations are
solutions of the classical equations of motion. 

  These solutions can be found using the identity
\be 
\label{Bog_ineq}
S  =  \frac{1}{4g^2}\int d^4x\, \left[\pm G^a_{\mu\nu} \tilde G^a_{\mu\nu}
 + \frac{1}{2} \left( G^a_{\mu\nu}\mp \tilde G^a_{\mu\nu}\right)^2
  \right],
\ee
where $\tilde G_{\mu\nu}=1/2\epsilon_{\mu\nu\rho\sigma}G_{\rho\sigma}$
is the dual field strength tensor (the field strength tensor in which
the roles of electric and magnetic fields are reversed). Since the 
first term is a topological invariant (see below) and the last term 
is always positive, it is clear that the action is minimal if the 
field is (anti) self-dual
\be
\label{self_dual}
G^a_{\mu \nu}=\pm\tilde G^a_{\mu \nu}.
\ee
The action of a self-dual field configuration is determined by its
topological charge
\be 
\label{top_charge}
Q =  {1\over 32\pi^2}\int d^4x\, G^a_{\mu\nu} \tilde G^a_{\mu\nu}.
\ee
From  (\ref{Bog_ineq}), we have $S=(8\pi^2|Q|)/g^2$. For finite action 
configurations, $Q$ has to be an integer. The instanton is a solution 
with $Q=1$ \cite{BPST}
\be  
\label{BPST_inst} 
A^a_\mu(x)= \frac{2\eta_{a\mu\nu}x_\nu}
  {x^2+\rho^2},
\ee 
where the 't Hooft symbol $\eta_{a\mu\nu}$ is defined by 
\be 
\label{eta_def}
\eta_{a\mu\nu} = \left\{ \begin{array}{rcl}
 \epsilon_{a\mu\nu}  \hspace{0.5cm}  \mu,\,\nu=1,\,2,\,3, \\
 \delta_{a\mu}                       \nu=4,  \\
-\delta_{a\nu}                       \mu=4.
\end{array}\right.
\ee
and $\rho$ is an arbitrary parameter characterizing the size of the 
instanton. This original instanton has its non-trivial topology at
large distances, but if we are to consider instanton ensemble, its another
form, the so called {\em singular gauge} on is needed 
\be   
A^a_\mu(x)= \frac{2\bar\eta_{a\mu\nu}x_\nu \rho^2}
  {(x^2+\rho^2)x^2},
\ee 
because in this case the non-trivial topology is at the point singularity.

The classical instanton solution has a number of degrees 
of freedom, known as collective coordinates. In addition to the
size, the solution is characterized by the instanton position $z_\mu$ 
and the color orientation matrix $R^{ab}$ (corresponding to color
rotations $A_\mu^a\to R^{ab}A_\mu^b$). A solution with topological
charge $Q=-1$ can be constructed by replacing $\eta_{a\mu\nu}\to
\overline\eta_{a\mu\nu}$, where $\overline\eta_{a\mu\nu}$ is defined
by changing the sign of the last two equations in (\ref{eta_def}).

   The physical meaning of the instanton solution becomes clear if
we consider the classical Yang-Mills Hamiltonian (in the temporal gauge,
$A_0=0$)
\be
\label{H_QCD}
 H = \frac{1}{2g^2}\int d^3x\,(E_i^2+B_i^2), 
\ee
where $E_i^2$ is the kinetic and $B_i^2$ the potential energy term. 
The classical vacua corresponds to configurations with zero field 
strength. For non-abelian gauge fields this limits the gauge fields 
to be ``pure gauge" $A_i= i U(\vec x) \partial_i U(\vec x)^\dagger$.
Such configurations are characterized by a topological winding number 
$n_W$ which distinguishes between gauge transformations $U$ that are 
not continuously connected.

  This means that there is an infinite set of classical vacua enumerated 
by an integer $n$. Instantons are tunneling solutions that connect
the different vacua. They have potential energy $B^2>0$ and kinetic
energy $E^2<0$, their sum being zero at any moment in time. Since 
the instanton action is finite, the barrier between the topological
vacua can be penetrated, and the true vacuum is a linear combination 
$|\theta\rangle =\sum_n e^{in\theta}|n\rangle$ called the theta vacuum. 
In QCD, the value of $\theta$ is an external parameter. If $\theta\neq 
0$ the QCD vacuum breaks CP invariance. Experimental limits on CP 
violation require\footnote{The question why $\theta$ happens 
to be so small is known as the ``strong CP problem". Most likely, the
resolution of the strong CP problem requires physics outside QCD and
we will not discuss it any further.} $\theta <10^{-9}$. 

  The rate of tunneling between different topological vacua is 
determined by the semi-classical (WKB) method. From the single
instanton action one expects
\be
P_{tunneling} \sim \exp(-8 \pi^2/g^2) .
\ee
The factor in front of the exponent can be determined by taking into
account fluctuations $A_\mu=A_\mu^{cl}+\delta A_\mu$ around the 
classical instanton solution. This calculation was performed in 
a classic paper by 't Hooft \cite{tHooft}. The result is
\be 
\label{eq_d(rho)} 
dn_I = \frac{0.47\exp(-1.68 N_c)}{(N_c-1)!(N_c-2)!}
 \left(\frac{8 \pi^2}{g^2}\right)^{2 N_c} 
 \exp\left(-\frac{8\pi^2}{g^2(\rho)}\right) 
 \frac{d^4zd\rho}{\rho^5},  
\ee
where $g^2(\rho)$ is the running coupling constant at the scale of
the instanton size. Taking into account quantum fluctuations, the
effective action depends on the instanton size. This is a sign of
the conformal (scale) anomaly in QCD. Using the one-loop beta function 
the result can be written as $ dn_I/(d^4z) \sim d\rho \rho^{-5}(\rho 
\Lambda)^b$ where $b=(11 N_c/3)=11$ is the first coefficient of the beta 
function. Since $b$ is a large number, small size instantons are strongly 
suppressed. On the other hand, there appears to be a divergence at large 
$\rho$. In this regime, however, the perturbative analysis based on the 
one loop beta function is not applicable. 

\subsection{\it Zero Modes and the $U(1)_A$ anomaly}
\label{sec_ua1}

  In the last section we showed that instantons interpolate between
different topological vacua in QCD. It is then natural to ask if the 
different vacua can be physically distinguished. This question is 
answered most easily in the presence of light fermions, because
the different vacua have different axial charge. This observation 
is the key element in understanding the mechanism of chiral anomalies.

   Anomalies first appeared in the context of perturbation theory
\cite{Adl_69,BJ_69}. From the triangle diagram involving an external 
axial vector current one finds that the flavor singlet current which 
is conserved on the classical level develops an anomalous divergence 
on the quantum level
\be 
\label{u1a_anom}
\partial_\mu j_\mu^5 = \frac{N_f}{16\pi^2} 
    G^a_{\mu\nu}\tilde G^a_{\mu\nu}.
\ee 
This anomaly plays an important role in QCD, because it explains 
the absence of a ninth goldstone boson, the so called $U(1)_A$ puzzle. 

  The mechanism of the anomaly is intimately connected with instantons. 
First, we recognize the integral of the RHS of (\ref{u1a_anom}) as 
$2N_f Q$, where $Q$ is the topological charge. This means that in 
the background field of an instanton we expect axial charge conservation
to be violated by $2N_f$ units. The crucial property of instantons, 
originally discovered by 't Hooft, is that the Dirac operator has a 
zero mode $iD\!\!\!\!/\,\psi_0(x)=0$ in the instanton field. For an 
instanton in the singular gauge, the zero mode wave function is
\be  
\label{eq_zm}
\psi_0(x)={\rho \over \pi} \frac{1}{(x^2+\rho^2)^{3/2}} 
 \frac{\gamma\cdot x}{\sqrt{x^2}}\frac{1+\gamma_5}{2} \phi 
\ee
where $\phi^{\alpha m}=\epsilon^{\alpha m}/\sqrt{2} $ is a constant spinor, 
which couples the color index $\alpha$ to the spin index $m=1,2$. Note 
that the solution is left handed, $\gamma_5 \psi_0=-\psi_0$. Analogously, 
in the field of an anti-instanton there is a right handed zero mode.

   We can now see how axial charge is violated during tunneling.
 For this purpose, let us consider the Dirac 
Hamiltonian $i\vec\alpha\cdot\vec D$ in the field of the instanton.
The presence of a 4-dimensional normalizable zero mode implies that 
there is one left handed state that crosses from positive to negative 
energy during the tunneling event. This can be seen as follows: In 
the adiabatic approximation, solutions of the Dirac equation are 
given by
\be
 \psi_i(\vec x,t) =  \psi_i(\vec x,t=-\infty)
  \exp\left(-\int_{-\infty}^{t}dt'\, \epsilon(t')\right).
\ee 
The only way we can have a 4-dimensional normalizable wave function
is if $\epsilon_i$ is positive for $t\to\infty$ and negative for $t\to
-\infty$. This explains how axial charge can be violated during 
tunneling. No fermion ever changes its chirality, all states simply
move one level up or down. The axial charge comes, so to say, from 
the ``bottom of the Dirac sea".

\subsection{\it The effective interaction between quarks}
\label{sec_leff}

   Proceeding from pure glue theory to QCD with light quarks, one has
to deal with the much more complicated problem of quark-induced 
interactions. Indeed, on the level of a single instanton we can 
not even understand the presence of instantons in full QCD. The
reason is again related to the existence of zero modes. In the 
presence of light quarks, the tunneling rate is proportional
to the fermion determinant, which is given by the product of the 
eigenvalues of the Dirac operator. This means that (as $m\to 0$) 
the tunneling amplitude vanishes and individual instantons cannot 
exist! 

   This result is related to the anomaly: During the tunneling event, 
the axial charge of the vacuum changes, so instantons have to be 
accompanied by fermions. The tunneling amplitude is non-zero only 
in the presence of external quark sources, because zero modes in 
the denominator of the quark propagator can cancel against zero modes 
in the determinant. Consider the fermion propagator in the instanton 
field
\be
\label{S_inst}
S(x,y) = \frac{\psi_0(x)\psi^+_0(y)}{im}
 +\sum_{\lambda\neq 0}\frac{\psi_\lambda(x)\psi^+_\lambda(y)}{\lambda+im} 
\ee
where $iD\!\!\!\!/\,\psi_\lambda=\lambda\psi_\lambda$. For $N_f$ light
quark flavors the instanton amplitude is proportional to $m^{N_f}$. 
Instead of the tunneling amplitude, let us calculate a $2N_f$-quark 
Green's function $\langle \prod_f \bar\psi_f(x_f)\Gamma \psi_f(y_f)
\rangle$, containing one quark and antiquark of each flavor.
Performing the contractions, the amplitude involves $N_f$ 
fermion propagators (\ref{S_inst}), so that the zero mode 
contribution involves a factor $m^{N_f}$ in the denominator.
 
    The result can be written in terms of an effective Lagrangian 
\cite{tHooft}. It is a non-local $2N_f$-fermion interaction, where  
the quarks are emitted or absorbed in zero mode wave functions. 
The result simplifies if we take the long wavelength limit (in
reality, the interaction is cut off at momenta $k>\rho^{-1}$) and
average over the instanton position and color orientation. For 
$N_f=1$ the result is \cite{tHooft,SVZ_80b}
\be 
\label{Leff_nf1}
{\cal L}_{N_f=1}= \int d\rho\, n_0(\rho) \left( m\rho- \frac{4}{3}
 \pi^2\rho^3 \bar q_R q_L \right), 
\ee
where $n_0(\rho)$ is the tunneling rate. Note that the zero mode
contribution acts like a mass term. For $N_f=1$, there is only one 
chiral $U(1)$ symmetry, which is anomalous. This means that the 
anomaly breaks chiral symmetry and gives a fermion mass term. This
is not true for more than one flavor. For $N_f=2$, the result is 
\be
\label{Leff_nf2}
{\cal L}_{N_f=2} = \int d\rho\, n_0(\rho) \bigg[  \prod_f
 \left( m\rho- \frac{4}{3}\pi^2\rho^3 \bar q_{f,R} q_{f,L} \right)
  \hspace{2cm} \ee
 $$  +\, \frac{3}{32}\left(\frac{4}{3}\pi^2\rho^3\right)^2
 \left( \bar u_R\lambda^a u_L \bar d_R\lambda^a d_L
  - \bar u_R\sigma_{\mu\nu}\lambda^a u_L 
    \bar d_R\sigma_{\mu\nu}\lambda^a d_L \right) \bigg]. \nonumber
$$
One can easily check that the interaction is $SU(2)\times SU(2)$
invariant, but $U(1)_A$ is explicitly broken. This Lagrangian is 
of the type first studied by Nambu and Jona-Lasinio \cite{NJL} 
and widely used as a model for chiral symmetry breaking and as an 
effective description for low energy chiral dynamics.
It can be transformed to the form discussed above when we compared it
to NJL interaction.
 
\subsection{\it The quark condensate in the mean field approximation}
\label{sec_qbarq}
     
  We showed in the last section that in the presence of light 
fermions, tunneling can only take place if the tunneling event 
is accompanied by $N_f$ fermions which change their chirality. 
But in the QCD vacuum, chiral symmetry is broken and the quark
condensate $\langle\bar qq\rangle =\langle\bar q_Lq_R+\bar q_Rq_L
\rangle$ is non-zero. This means that there is a finite amplitude 
for a quark to change its chirality and we expect the instanton
density to be finite. 

  For a sufficiently dilute system of instantons, we can estimate 
the instanton density in full QCD from the expectation value of
the $2N_f$ fermion operator in the effective Lagrangian 
(\ref{Leff_nf2}). Using the factorization assumption \cite{SVZ}, 
we find that the factor $\prod_f m_f$ in the instanton density should 
be replaced by $\prod_f m^*_f$, where the effective quark mass is 
given by
\be  
\label{eff_mass}
m_f^*  =  m_f- {2\over 3}\pi^2\rho^2 \langle \bar q_f q_f\rangle  .
\ee  
This shows that if chiral symmetry is broken, the instanton density 
is finite in the chiral limit. 

  This obviously raises the question whether the quark condensate itself 
can be generated by instantons. This question can be addressed using 
several different techniques (for a review, see \cite{SS_98,Diakonov}). 
One possibility is to use the effective interaction (\ref{Leff_nf2})
and to calculate the quark condensate in the mean field (Hartree-Fock)
approximation. This correspond to summing the contribution of all 
``cactus" diagrams to the full quark propagator. The result is a 
gap equation \cite{DP_86}
\be 
\label{gap}
\int \frac{d^4k}{(2\pi)^4} \frac{M^2(k)}{k^2+M^2(k)}  = 
\frac{N}{4N_cV},
\ee 
which determines the constituent quark mass $M(0)$ in terms of 
the instanton density $(N/V)$. Here, $M(k)=M(0)k^2\varphi^{\prime 2}
(k)/(2\pi\rho)$ is the momentum dependent effective quark mass and
$\varphi^{\prime}(k)$ is the Fourier transform of the zero 
mode profile \cite{DP_86}. The quark condensate is given by
\be
\label{qbarq_hfa}
\langle \bar qq\rangle   =  
 -4N_c \int\frac{d^4k}{(2\pi)^4} \frac{M(k)}{M^2(k)+k^2}.
\ee
Using our standard parameters $(N/V)=1\,{\rm fm}^{-4}$ and $\rho=
1/3$ fm, one finds $\langle\bar qq\rangle \simeq -(255\,{\rm 
MeV})^3$ and $M(0)=320$ MeV. Parametrically, $\langle\bar qq
\rangle \sim (N/V)^{1/2}\rho^{-1}$ and $M(0)\sim (N/V)^{1/2}
\rho$. Note that both quantities are not proportional to $(N/V)$,
but to $(N/V)^{1/2}$. This is a reflection of the fact that 
spontaneous breaking of chiral symmetry is not a single instanton
effect, but involves infinitely many instantons. 

  A very instructive way to study the mechanism for chiral symmetry 
breaking at a more microscopic level is by considering the distribution
of eigenvalues of the Dirac operator. A general relations that connects 
the spectral density $\rho(\lambda)$ of the Dirac operator to the quark 
condensate was given by Banks-Casher relation
\be
\label{BC}
\langle\bar qq\rangle  =  -\pi \rho(0). 
\ee
This result is analogous to the Kondo formula for the electrical
conductivity. Just like the conductivity is given by the 
density of states at the Fermi surface, the quark condensate 
is determined by the level density at zero virtuality $\lambda$. 
For a disordered, random, system of instantons the zero modes 
interact and form a band around $\lambda=0$. As a result, the 
eigenstates are de-localized and chiral symmetry is broken. On 
the other hand, if instantons are strongly correlated, for 
example bound into topologically neutral molecules, the eigenvalues
are pushed away from zero, the eigenstates are localized and chiral
 symmetry is unbroken. As we will see below, 
precisely which scenario is realized depends on the parameters
of the theory, like the number of light flavors and the temperature. 
Of course, for ``real" QCD with two light flavors at $T=0$, we 
expect chiral symmetry to be broken. This is supported by numerical
simulations of the partition function of the instanton liquid, see
\cite{SS_98}.

\subsection{The Qualitative Picture of the Instanton Ensemble}
 Using basically such expressions and the known value of
the quark condensate it was
 pointed out  in\cite{Shu_82} that all would be consistent only if
  the typical instanton size happened to be significantly smaller than their
 separation\footnote{Derived in turn from the gluon condensate and the
topological susceptibility.}, $R=n^{-1/4}\approx 1 fm$, namely
$
\rho_{\rm max}\sim 1/3\,{\rm fm} .
$

\begin{figure}[h]
\vskip .05in
 \centering
 \includegraphics[width=3.5in]{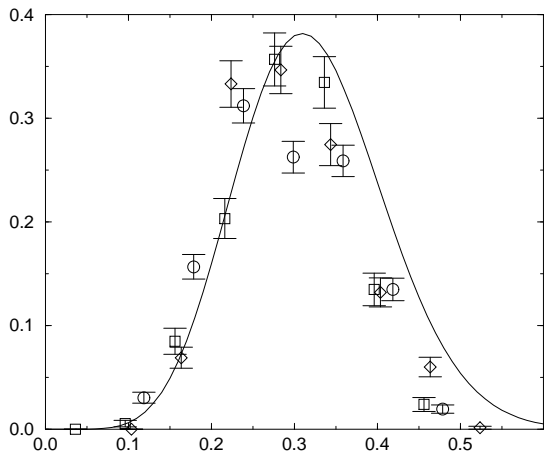}
  \vspace{-.05in}
  \caption{\label{fig_sizes}
 The instanton density $dn/d\rho d^4z$, [fm$^{-5}$] versus its size
 $\rho$ [fm]. 
 The points are from the lattice work \protect\cite{anna},
for this theory, with 
$\beta$=5.85 (diamonds), 6.0 (squares) and 6.1 (circles). Their
comparison should demonstrate that results are
rather lattice-independent.
The line corresponds to the proposed
expression $\sim exp(-2\pi\sigma\rho^2)$, see text.
  }
\end{figure}

 In Fig.(\ref{fig_sizes}) one can see  lattice data on instanton size
distribution, obtain by cooling of the original gauge fields.
Similar distribution can also be obtained from fermionic lowest Dirac
eigenmodes: in this case no ``cooling'' is needed.

\begin{figure}[h]
\vskip .05in
 \centering
 \includegraphics[width=2.5in, angle=270]{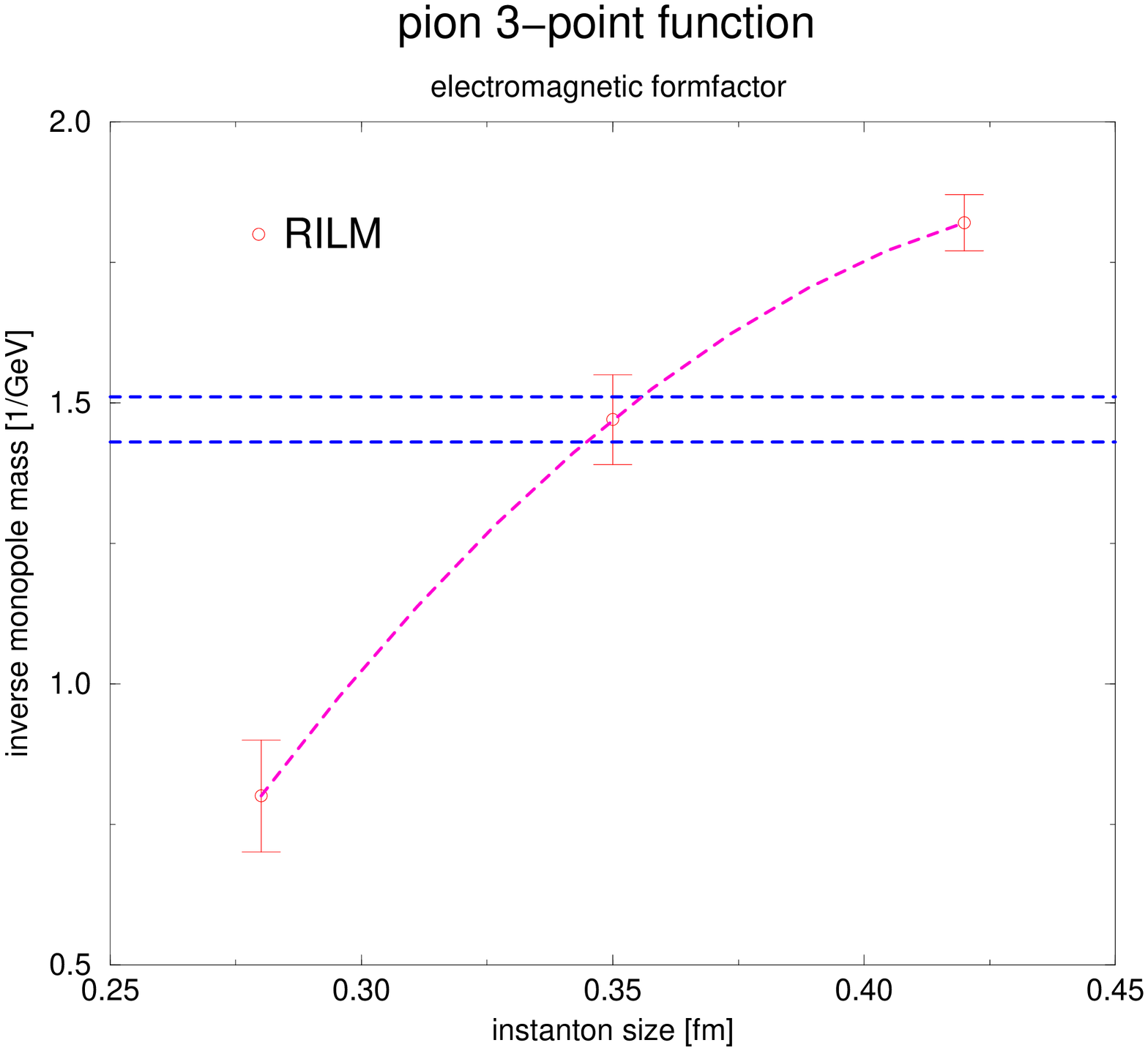}
 \vskip 0.1in 
  \caption{\label{fig_pionff}
The fitted parameter M of the pion form-factor $ff\sim M^2/(Q^2+M^2)$
versus the inputed instanton size.  }
\end{figure}

Let me now show another evidence
for this value of the instanton size,  taken from the pion form-factor
calculated\cite{BS} in the instanton model. In Fig.(\ref{fig_pionff})
we show how the experimentally measured pion size correlates with the
input mean instanton size: one can see from it
that the value .35 fm is a clear winner.

If so,  the following qualitative
picture of the QCD vacuum have emerged:
\begin{enumerate}

\item Since the instanton size is significantly smaller than the typical 
separation $R$ between instantons, $\rho/R \sim 1/3$, the vacuum is fairly 
dilute. The fraction of spacetime occupied by strong fields is only a few 
percent.

\item The fields inside the instanton are very strong, $G_{\mu\nu}\gg\Lambda_{QCD}^2$.
This means that the semi-classical approximation is valid, and the typical 
action is large
\be
\label{S_typ}
 S_0 = 8\pi^2/g^2(\rho) \sim 10-15 \gg 1 .
\ee
Higher order corrections are proportional to $1/S_0$ and presumably small.

\item Instantons retain their individuality and are not destroyed by 
interactions. From the dipole formula, one can estimate
\be
\label{S_int_typ}
 |\delta S_{int}| \sim (2-3) \ll S_0  .
\ee

\item Nevertheless, interactions are important for the structure of the 
instanton ensemble, since
\be
  \exp|\delta S_{int}| \sim 20 \gg 1 .
\ee
This implies that interactions have a significant effect on
correlations 
among 
instantons, and the instanton ensemble in QCD is not a dilute gas but an 
  {\em interacting liquid}. 
\end{enumerate}

 The aspects of the QCD vacuum for which  instantons are most
 important are
those related to
light fermions.
Their importance  in the context of chiral symmetry breaking is 
related to the fact that the Dirac operator has a chiral zero mode in the 
field of an instanton. 
These zero modes are  localized quark states around instantons, like
atomic states of electrons around nuclei. At finite density of  instantons
those states can become collective, like atomic states in metals. 
The resulting de-localized state corresponds to the wave function of the quark
condensate. 

Direct tests of all these ideas on the lattice are possible.
One may have a look at the lowest eigenmodes and see if they
are related to instantons or something else (monopoles, vortices...)
by identifying their shapes - 4d bumps (lines or 2-d sheets)
respectively. So far, only bumps (that is the instantons) were seen.

One may also test how locally chiral are the lowest eigenmodes. Just
at this school I learned from one of its young participants,
Christof Gattringer, about his version of chiral lattice fermions and
nice results he got. Among those is remarkably well defined separation 
between instanton-based and perturbative-like lowest modes, revealed
by
the so called participation ratios. 

Let me now explain about the {\em lowest QCD scale} generated by
instantons,
mentioned above. The width of the {\em zero mode zone} of states is of 
the order of root-mean-square matrix element of the Dirac operator
$<I|\slash D |J>\sim \rho^2/R^3$. Here states I,J are some instanton
and anti-instanton zero modes, rho is the instanton size and $R\sim
n^{-1/4}\approx 1 fm$ is the
distance between their centers. Note small factor $(\rho/R)^2\sim 1/10$
here. The Dirac eigenvalues from the zone have similar magnitude. Now,
the eigenvalues enter together with quark mass m: and so only when
this quark mass is smaller than this scale we start seeing the physics 
of the zero mode zone. In particular, for {\em quenched} QCD (or
instanton liquid) there is no determinant and the zone states have
rather wrong spectrum. However, only if the quark mass is small
compared
to its width we start observing the difference. Only recently lattice
practitioners were able to do so: indeed, quenched QCD results at
small m start deviating from the correct answers quite drastically.

\subsection{Interacting instantons}
In the QCD partition function there are two types of fields, gluons and quarks,
and so the first question one addresses is {\em which integral to take first}.

(i) One way is to eliminate $gluonic$ degrees of freedom first. Physical 
motivation for this may be that gluonic states are heavy and an effective 
fermionic theory should be better suited to derive an effective low-energy
fermionic theory. 
It is a well-trodden path and one can follow it to the development of 
a similar four-fermion theory, the NJL model. One can do simple mean field or 
random field approximation (RPA) diagrams, and find the mean condensate and 
properties of the Goldstone mesons\cite{DP_86}. The results for
 Color Super-conductors at high density reported below are done 
with the same technique as well. 
But nevertheless, not much can really be done in such NJL-like approach.
In fact, multiple attacks during the last 40 years at the NJL model
{\em beyond the mean field} 
 basically failed. In particular, one might think that since
baryons are states with three quarks, and
one may wonder if  using quasi-local 
four-fermion Lagrangians for the three body problem is 
a solvable quantum mechanical problem,
and one can at least tell if nucleons are or are not bound in NJL.
 In fact it is not: the results 
depend strongly on subtleties of how the local limit for the interaction
is defined, and there is no clear answer to this question.
Other notorious
attempts to sum more complicated diagrams deal with
the possible modification
of the the chiral condensate. Some works even claim that
those diagrams destroy it $completely$!

Going from NJL to instantons improves the situation enormously:
the shape of the form-factor is no longer a guess (it is provided by the
shape of zero modes) and one can in principle evaluate any particular diagram.
However {\em summing them all up} still seems like an impossible task.

(ii) The solution to this problem was found. For that
 one has to follow the opposite strategy and do the $fermion$
 integral first. The first
step  is simple and standard: fermions
 only enter quadratically, leading to a fermionic determinant.
 In the instanton approximation, it leads
to the Interacting Instanton Liquid Model, defined by the following partition 
function:
\be 
\label{part_fct} 
Z =   \sum_{N_+,\, N_-} {1 \over N_+ ! N_- !}\int 
    \prod_i^{N_+ + N_-} [d\Omega_i\; d(\rho_i) ] 
    \exp(-S_{\rm int})\prod_f^{N_f} \det(\hat D+m_f) \, , 
\ee 
describing a system of pseudo-particles interacting via the bosonic action and 
the fermionic determinant. Here $d\Omega_i=dU_i\, d^4z_i\, d\rho_i$ is the 
measure in color orientation, position and size associated with single 
instantons, and  $d(\rho)$ is the single instanton density 
$d(\rho)= dn_{I,\bar I}/d\rho dz$.
 
The gauge interaction between instantons is approximated by a sum of pure 
two-body interaction $S_{\rm int}=\frac{1}{2}\sum_{I\neq J}S_{\rm int} 
(\Omega_{IJ})$. Genuine three body effects in the instanton interaction are not
important as long as the ensemble is reasonably dilute. Implementation of this 
part of the interaction (quenched simulation) is quite analogous to usual 
statistical ensembles made of atoms. 
 
As already mentioned, quark exchanges between instantons are included in the 
fermionic determinant. Finding a diagonal set of fermionic eigenstates  of the 
Dirac operator is similar to what people are doing, e.g., in quantum chemistry
when electron states for molecules are calculated. The difficulty of our 
problem is however much higher, because this set of fermionic states should be
determined for $all$ configurations which appear during the Monte-Carlo process.

If the set of fermionic states is however limited to the subspace of instanton 
zero modes, the problem becomes tractable numerically. Typical calculations in 
the IILM involved up to N$\sim 100$ instantons (+anti-instantons):
 which means that 
the determinants of $N\times N$ matrices are involved. Such determinants can be 
evaluated by an
 ordinary workstation (and even PC these days) so quickly that a 
straightforward Monte Carlo simulation of the IILM is possible in a matter of 
minutes. On the other hand, expanding the determinant in a sum of products of 
matrix elements, one can 
easily identify the sum of all closed loop diagrams up 
to order $N$ in the 't Hooft interaction.
 Thus, in this way one can actually take care of 
about 100 factorial diagrams!
 
\section{  Hadronic Structure   and the  QCD correlation functions.}

\subsection{ Correlators as a bridge between { hadronic} and
{partonic} worlds}
 Consider two currents separated by a $space-like$ distance $x$
(which can be considered as
the spatial distance, or an Euclidean time) and introduce  correlation functions of the type
\be 
K(x)=<T(J(x) J(0))> 
\ee
with $J(x)=\bar \psi(x)  \Gamma \psi(x)$. The matrix
$\Gamma$ contains $\gamma_\mu$ for vector currents,
$\gamma_5$ for the pseudoscalar or $1$ for
the scalars, etc,
 and also a flavor matrix, if needed.

We will start with
isovector vector and axial currents, and then
 discuss  4 scalar-pseudoscalar  channels: $\pi$ (P=-1,
I=1), $\sigma$ or $f_0$ (P=+1,I=0), $\eta$ (P=-1,I=0) and 
$\delta$ or $a_0$ (P=+1,I=1). 

 In a (relativistic) field theory, correlation functions of gauge 
invariant local operators are the proper tool to study the spectrum 
of the theory. The correlation functions can be calculated either
from the physical states (mesons, baryons, glueballs) or in terms
of the fundamental fields (quarks and gluons) of the theory. In the
latter case, we have a variety of techniques at our disposal, ranging
from perturbative QCD, the operator product expansion (OPE), to
models of QCD and lattice simulations. For this reason, correlation
functions provide a bridge between hadronic phenomenology on the one
side and the underlying structure of the QCD vacuum on the other side.

   Loosely speaking, hadronic correlation functions play the same role 
for understanding the forces between quarks as the $NN$ scattering 
phase shifts did in the case of nuclear forces. In the case of quarks, 
however, confinement implies that we cannot define scattering amplitudes
in the usual way. Instead, one has to focus on the behavior of gauge 
invariant correlation functions at short and intermediate distance 
scales. The available theoretical and phenomenological information 
about these functions was recently reviewed in \cite{Shu_93}.

In all cases at small x we expect
$K(x)\approx K_0(x)$ where the latter corresponds to
just $free$ propagation of (about massless) light quarks.
The zeroth order correlators are all just $K_0(x)=12/(\pi^4 x^{6})$,
basically the square of the massless
quark propagator. 

The first deviations due to non-perturbative effects can be studied 
using Wilsonian Operator Product Expansion (OPE) in ref\cite{SVZ}.
For all scalar and pseudoscalar channels
the resulting first correction is
\be
{K(x)\over K_0(x)}=1+ {x^4\over 384} <(gG)^2>+...
\ee
The ``gluon condensate"  is assumed to be  made out of a soft
 vacuum field, 
and therefore all  arguments can be simply taken at the point $x=0$.
 The so-called $standard$ value of the ``gluon condensate" appearing here
was estimated previously from charmonium sum rules:
\be
<(gG)^2>_{SVZ}\approx .5 \, GeV^4
\ee
Thus, the OPE suggests the following scale, at which the correction becomes
equal to the first term:
\be
x_{OPE}=(384/<(gG)^2>_{SVZ})^{1/4}\approx 1.0 \, \, fm
\ee
This seems to be completely consistent with the  approximation used.
  However, as  Novikov, Shifman, Vainshtein and Zakharov
soon noticed\cite{NSVZ}, this (and other OPE corrections) 
completely failed to describe all the  $J^{P}=O^{\pm}$ channels:
we return to this issue after we consider vectors and axials.

\subsection{Vector and axial correlators}

  The information available on vector correlation functions 
 from experimental data on $e^+e^-->hadrons$, 
the OPE and other exact results was 
reviewed in \cite{Shu_93}. Since then, however, new  high statistics 
measurement of hadronic $\tau$ decays $\tau\rightarrow \nu_\tau
+{\rm hadrons}$ have been done. For definiteness, we use results of
one of them, ALEPH experiment at CERN \cite{aleph1,aleph2}. 

\begin{figure}[t]
\begin{center}
\includegraphics[width=7cm]{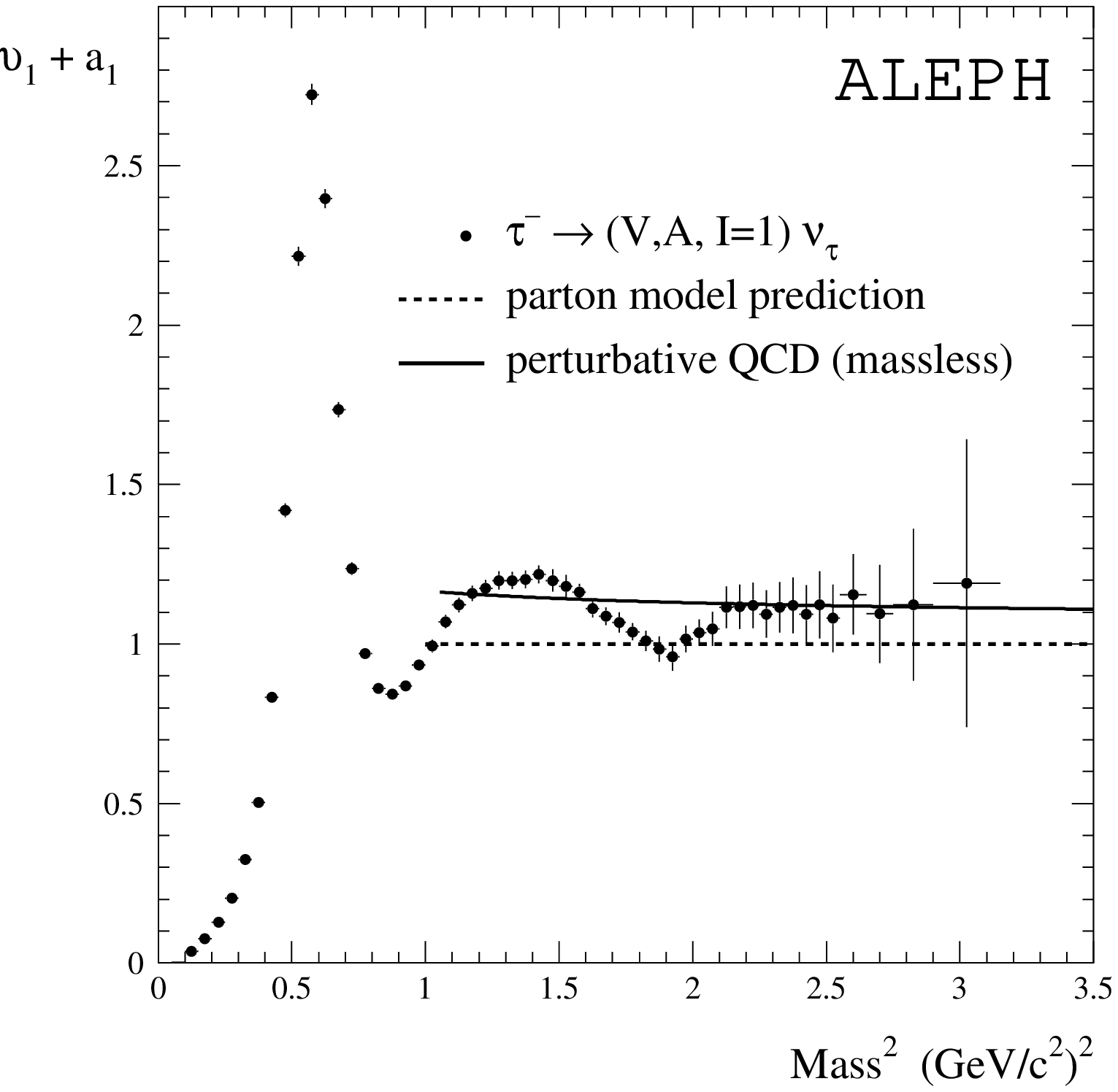}
\includegraphics[width=7cm]{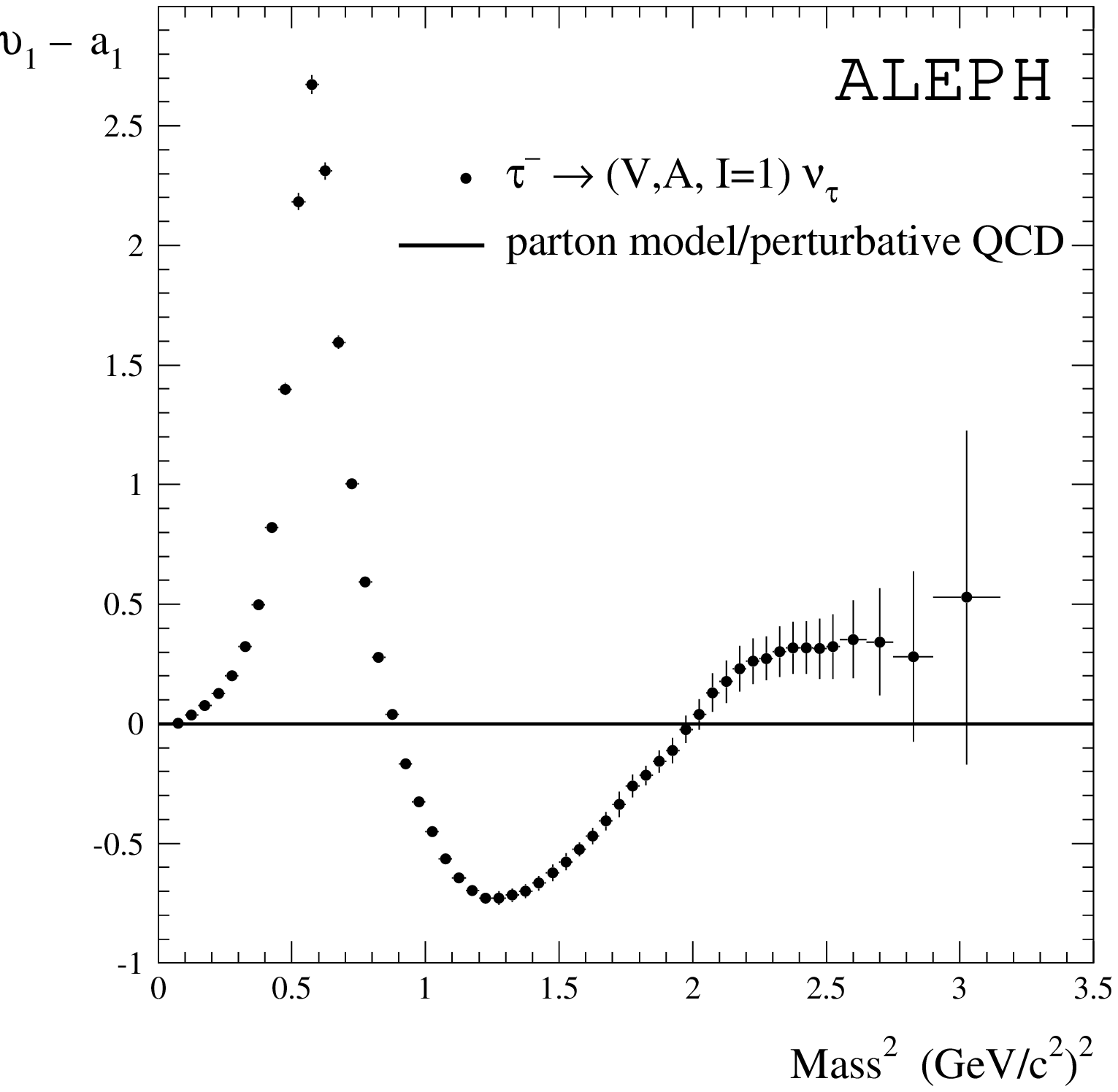}
\caption{\label{fig_data}
Spectral functions $v(s)\pm a(s)=4\pi^2(\rho_V(s)+\rho_A(s))$
extracted by the ALEPH collaboration from $tau$ lepton hadronic decays.}
\end{center}
\end{figure}

 The vector and 
axial-vector correlation functions  are $\Pi_V(x) = \langle 
j^a_\mu(x)j^a_\mu(0)\rangle$ and $\Pi_A(x)=\langle
j^{5\, a}_\mu(x) j^{5\, a}_\mu(0)\rangle$. Here, $j^a_\mu(x)
=\bar{q}\gamma_\mu\frac{\tau^a}{2}q$, $j^{5\,a}_\mu(x)
=\bar{q}\gamma_\mu\gamma_5\frac{\tau^a}{2}q$ are the 
isotriplet vector and axial-vector currents. The Euclidean
correlation functions have the spectral representation \cite{Shu_93}
\be
 \Pi_{V,A}(x) = \int ds\, \rho_{V,A}(s)D(\sqrt{s},x),
\ee
where $D(m,x) = m/(4\pi^2 x)K_1(mx)$ is the Euclidean 
coordinate space propagator of a scalar particle with 
mass $m$. We shall focus on the linear combinations
$\Pi_V+\Pi_A$ and $\Pi_V-\Pi_A$. These combinations 
allows for a clearer separation of different 
non-perturbative effects. The corresponding spectral
functions $\rho_V\pm\rho_A$ measured by the ALEPH
collaboration are shown in Fig. \ref{fig_data}. 
The errors are a combination of statistical and systematic ones (below we use 
them conservatively, as pure systematic): the main problem seems to be
separation into V and A of channels with Kaons, which may affect $V-A$ 
at $s>2 \, GeV$ at 10\%  level. None of our conclusions are sensitive
to it.

%

\begin{figure}[h]
\begin{center}
\leavevmode
\includegraphics[width=8cm,angle=-90]{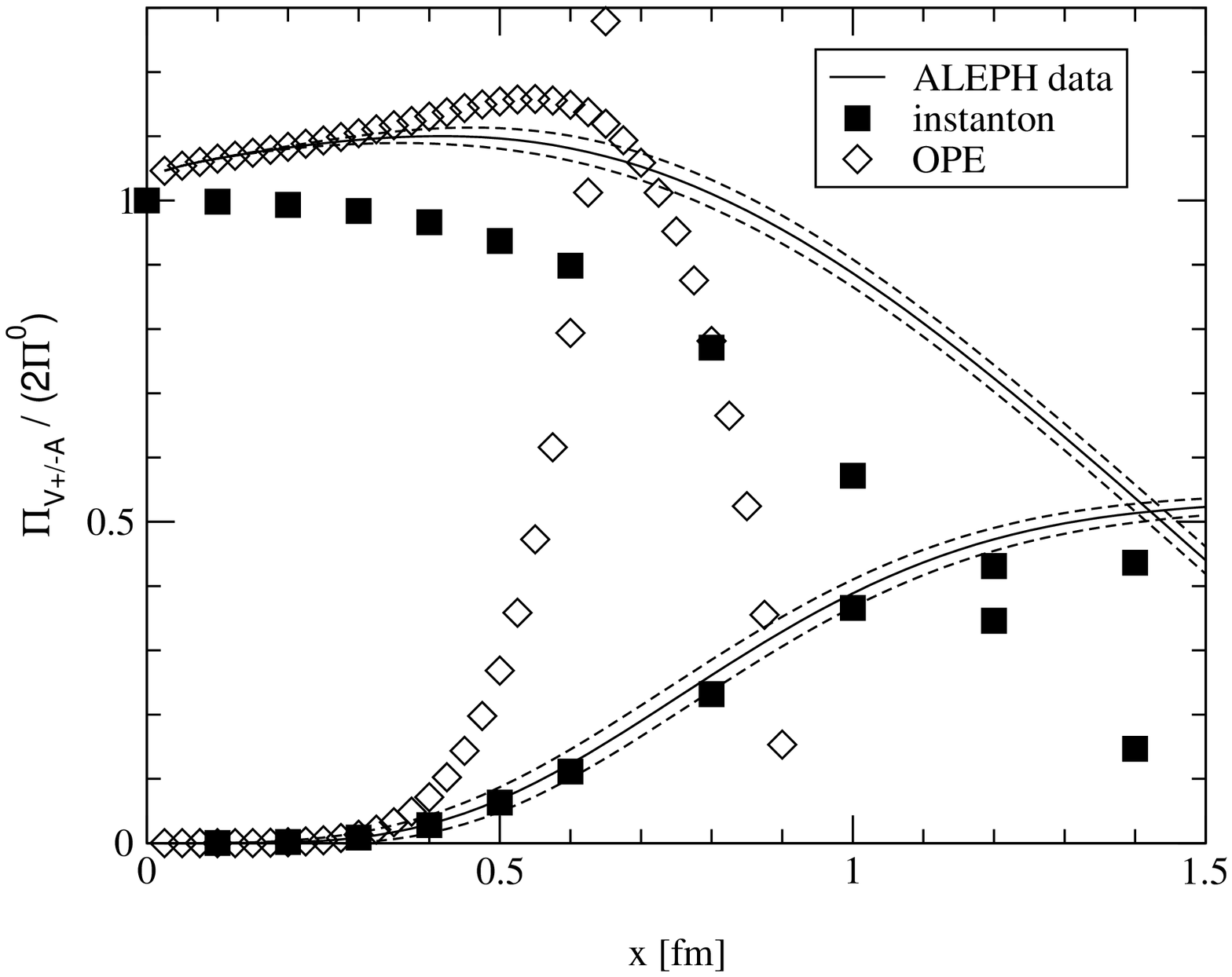}
\end{center}
\vspace*{-0.5cm}
\caption{\label{fig_cor1}
Euclidean coordinate space correlation functions $\Pi_V(x)
\pm \Pi_A(x)$ normalized to free field behavior. The solid lines
show the correlation functions reconstructed from the ALEPH spectral
functions and the dotted lines are the corresponding error band.
The squares show the result of a random instanton liquid model
and the diamonds the OPE fit described in the text.}
\end{figure}

 In QCD, the vector and axial-vector spectral functions must
satisfy chiral sum rules. Assuming that $\rho_V-\rho_A=0$ at  above
$s>m^2_\tau$, and using ALEPH data below it, one finds that all 4 
of the sum rules are satisfied within the experimental uncertainty,
but the central 
values  differ significantly from the chiral 
predictions \cite{aleph1}. In general, both functions are expected to
have oscillations of decreasing amplitude, and putting
$\rho_V-\rho_A$ to zero at arbitrary point imply appearance of  spurious dimension $d=2,4$
operators in the correlation functions at small x.
Therefore, we have decided to terminate the data above a specially tuned point,
 $s_0=2.5\,{\rm GeV}^2$,
 enforcing all 4  chiral sum rules. (The reader should
however be aware of the fact that we have, in effect, 
slightly moved the data points in the small $x$ region
within the error band.) Finally we add the pion pole 
contribution (not shown in Fig. \ref{fig_data}), which 
 corresponds to 
an extra term $\Pi_A^\pi(x) = f_\pi^2m_\pi^2 D(m_\pi,x)$. The resulting 
correlation functions $\Pi_V(x)\pm\Pi_A(x)$ are shown 
in Figs. \ref{fig_cor1}.

 We begin our analysis with the combination $\Pi_V-\Pi_A$. This 
combination is sensitive to chiral symmetry breaking, while 
perturbative diagrams, as well as gluonic operators cancel out.

 In Fig. \ref{fig_cor1} we compare the measured correlation
functions with predictions from the instanton liquid model
(in its simplest form, random instanton liquid with parameters n,
$\rho$ fixed in \cite{Shu_82} and
discussed above).

 The agreement of the instanton prediction with the measured
$V-A$ correlation is impressive: it extends all the way from
short to large distances. At distances $x>1.25$ fm both
combinations are dominated by the pion contribution
while at intermediate $x$ the $\rho,\rho'$ and $a_1$
resonances contribute.

 We shall now focus our attention on the $V+A$ correlation function.
The unique feature of this function is the fact that the correlator
remains close to free field behavior for distances as large 
as 1 fm. This phenomenon was referred to as ``super-duality''
in \cite{Shu_93}. 
 The instanton model reproduces this feature of the $V+A$ correlator.
We also notice that for small $x$ the deviation of the correlator
in the instanton model from free field behavior is small compared
to the perturbative $O(\alpha_2/\pi)$ correction. This opens the 
possibility of precision studies of the pQCD contribution. But 
before we do so, let us compare the correlation functions to
the OPE prediction
\be
\frac{\Pi_V(x)+\Pi_A(x)}{2\Pi_0(x)} &=&  1+\frac{\alpha_s}{\pi}
  - \frac{1}{384}\langle g^2G_{\mu\nu}^2 \rangle x^4 \nonumber \\
 & & \hspace{0.5cm}
  -\frac{4\pi^3}{81} \alpha_s(x)\langle{\bar{q}q}\rangle
   \log(x^2)x^6+\ldots 
\ee
Note that the perturbative correction is attractive, while the
power corrections of dimension $d=4$ and $d=6$ are repulsive.
Direct instantons also induce an $O(x^4)$ correction
$ 1 - \frac{\pi^2}{12} \left(\frac{N}{V}\right) x^4 + \ldots$ 
,
which is consistent with the OPE because in a dilute 
instanton liquid we have $\langle g^2G^2\rangle = 32\pi^2(N/V)$. 
This term can indeed be seen in the instanton calculation
and causes the correlator to drop below 1 at small $x$.
 It is possible to extract the value of $\Lambda_{QCD}$  (we find
$\alpha_s(m_\tau)=0.35$) and even clear indication of running coupling.
It is only possible to do because
the non-perturbative corrections (represented by instantons)
are basically cancelling each other to very high degree, in V+A
channel.

Why is it happening?
The first order in 't Hooft is indeed absent, due to chirality mismatch.
There is no general theoretical reason why all non-perturbative of higher
order should also do so: but ALEPH data used wrongly hint that they
actually do so.

\subsection{Spin-zero correlation functions}

Now we will see case4s which are completely opposite to
those just considered: the instanton-induced effects would be  large.
Furthermore, the 4 channels
actually show completely different non-perturbative 
deviation from $K_0$ at small x: half of them ($\pi,\sigma$)
deviate upward, and another pair ($\eta,\delta$) deviate downward.

But let me first demonstrate that the OPE scale determined above
cannot be right. All we have to do is to
 evaluate the strength of the  pion contribution 
to the correlator in question:
\be
K_\pi(x) = {\lambda_\pi^2\over 4 \pi^2 x^2} 
\ee
The coupling constant is defined as $ \lambda_\pi=<0|J(0)|\pi>$ and the
rest is nothing more than the scalar massless propagator\footnote{We can ignore
the pion mass at the
 distances in question. We also ignore contributions of other
states, which can only add positively to the correlator and made
disagreement only worse.}.
Because both the pion term and the gluon condensate correction
happen to be $1/x^2$, let us compare the coefficients. Ideal matching
would mean they are about the same
\be
\lambda_\pi^2 \approx {<(gG)^2>_{SVZ} \over 8 \pi^2}
\ee
The r.h.s. is about 0.0063 GeV$^4$. However, phenomenology tells us that 
 (unlike the better known coupling to
the axial current $f_\pi$) the coupling $ \lambda_\pi$ is surprisingly
large\footnote{The reason for that is the the pion is rather compact and also
the  wave function is concentrated at its center, so that 
its value at $r=0$ is large. We return to this point in the discussion of
the ``instanton liquid" model.}. The l.h.s. of this relation is actually
 $ \lambda_\pi^2=(.48\, GeV)^4=0.053\, GeV^4$,  about {\em 10 times larger}
than the r.h.s. It means  much larger non-perturbative effect
is needed to explain the deviation from the perturbative behavior.

Now, let us see why is it so.
The
instanton effects in spin-0 
channels are in these cases much larger because effect of 't
Hooft
interaction appears in those cases in the first order.
Furthermore, since it its flavor structure is non-diagonal $(\bar u u
)(\bar d d)$
 the correlator of two $\pi^0$ currents 
$(\bar u\gamma_5 u-\bar d\gamma_5 d)$
 have it with opposite sign as compared to
the correlator of $\eta'$ currents $(\bar u\gamma_5 u+\bar d\gamma_5
d)$.
What it means is that instantons are as attractive in the pion channel 
as they are repulsive in the $\eta'$ case. The situation is reversed
in the scalar channels: the isoscalar sigma is attractive and isovector 
is repulsive.

\begin{figure}
\vspace{-3mm}
\centerline{\includegraphics[width=6cm]{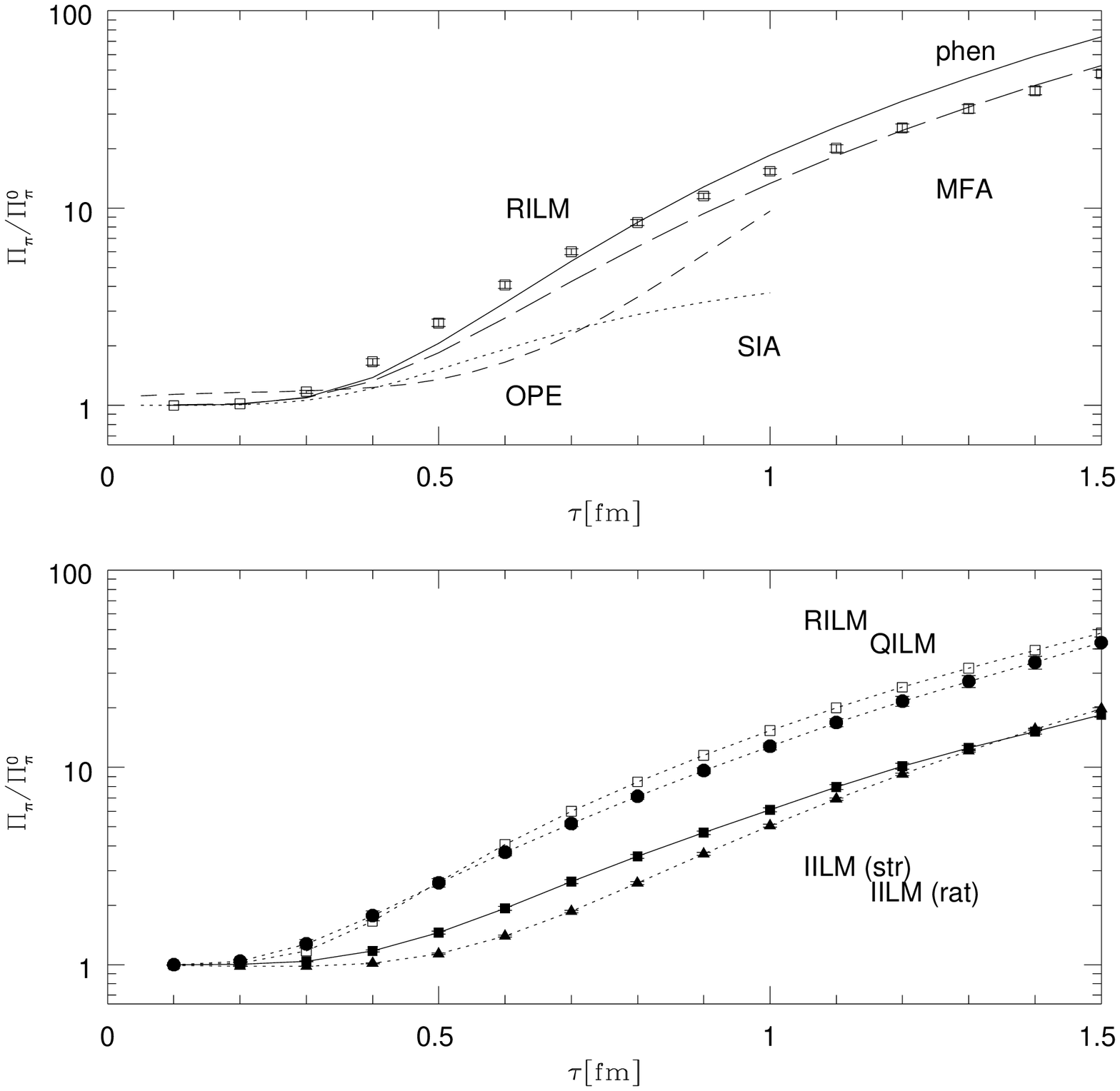}}
\vspace{-3mm}
\caption{\label{fig_pi_cor}
Pion correlation function in various approximations and instanton ensembles. In
the top figure we show the phenomenological expectation (solid), the OPE 
(dashed), the single instanton (dash-dotted) and mean field approximations 
(dashed) as well as data in the random instanton ensemble. In the bottom figure
we compare different instanton ensembles, random (open squares), quenched 
(circles) and interacting (streamline: solid squares, ratio ansatz solid 
triangles).}
\end{figure}

  Full results from versions of the
 instanton liquid model for pion correlators are
shown in fig.\ref{fig_pi_cor}.
 Different versions of the model (mentioned in figures below as
IILM(rat) etc) differ by a particular ansatz for the gauge field used,
from which the interaction is calculated. Note also, that these figures
 contain also a curve marked ``phen'': this is what the
correlator actually looks like, according to phenomenology.

\begin{figure}[t!]
\includegraphics[width=8cm]{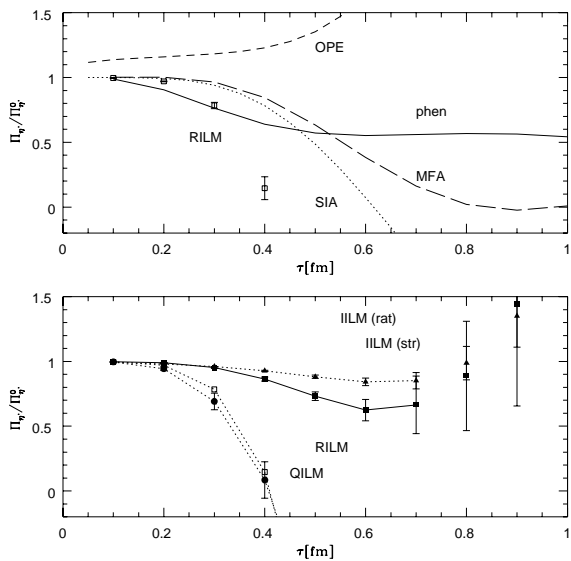}
\vspace{-3mm}
\caption{\label{fig_eta_cor}
Eta prime meson correlation functions. The various curves and data sets are
labeled as in Fig.~\ref{fig_pi_cor}. Note that random instanton liquid 
model (RILM) and quenched version (no fermionic determinant, only
bosonic
interactions) predict $\eta'$ correlator to go negative. The same
unphysical behavior has been found on the lattice.  }
\end{figure}

We simply  show a few results of 
correlation functions in the different instanton ensembles
(see original refs in\cite{SS_98}).
Some of them (like vector and axial-vector ones) turned out to be
easy:
nearly any variant of the
instanton model can reproduce  the (experimentally
known!)
correlators well.
 Some of them are sensitive to details of the model very much:
two such cases are shown in Figs.~\ref{fig_pi_cor}-\ref{fig_eta_cor}. The pion 
correlation functions in the different ensembles are qualitatively very similar.
The differences are mostly due to different values of the quark condensate 
(and the physical quark mass) in the different ensembles. Using the
 Gell-Mann-Oaks-Renner relation, one can extrapolate the pion mass to the physical value 
of the quark masses. The results are consistent with the experimental value in 
the streamline ensemble (both quenched and unquenched), but clearly too small 
in the ratio ansatz ensemble. This is a reflection of the fact that the ratio 
ansatz ensemble is not sufficiently dilute.

The situation is drastically different in the $\eta'$ channel. Among the $\sim 
40$ correlation functions calculated in the random ensemble, only the $\eta'$ 
and the isovector-scalar $\delta$ were found to be completely
unacceptable. The correlation function decreases very rapidly and becomes 
$negative$ at $x\sim 0.4$ fm. This behavior is incompatible even with a normal 
spectral representation. The interaction in the random ensemble is too 
repulsive, and the model ``over-explains" the $U(1)_A$ anomaly. 

The results in the unquenched ensembles (closed and open points) significantly 
improve the situation. This is related to dynamical correlations between 
instantons and anti-instantons (topological charge screening). The single 
instanton contribution is repulsive, but the contribution from pairs is 
attractive. Only if correlations among instantons and 
anti-instantons are sufficiently strong are the correlators  prevented from 
becoming negative. Quantitatively, the $\delta$ and $\eta_{\rm ns}$ masses in 
the streamline ensemble are still too heavy as compared to their experimental 
values. In the ratio ansatz, on the other hand, the correlation functions even 
show an enhancement at distances on the order of 1 fm, and the fitted masses 
are too light. This shows that the $\eta'$ channel is very sensitive to the 
strength of correlations among instantons.

In summary, pion properties are mostly sensitive to global properties of the 
instanton ensemble, in particular its diluteness. Good phenomenology demands 
$\bar\rho^4 n\simeq 0.03$, as originally suggested in\cite{Shu_82}. The 
properties of the $\rho$ meson are essentially independent of the diluteness, 
but show  sensitivity to $\bar I I$ correlations. These correlations become 
crucial in the $\eta'$ channel.  

\subsection{Baryonic correlation functions}
\label{sec_bar_cor}

 The existence of a 
strongly attractive interaction in the pseudoscalar quark-antiquark (pion) 
channel also implies an attractive interaction in the scalar quark-quark 
(diquark) channel. This interaction is phenomenologically very desirable, 
because it immediately explains why the nucleon  is light, while 
the delta (S=3/2,I=3/2) is heavy.

The so called  Ioffe currents (with no derivatives
and the minimum number of quark fields) are local operators
which can excite states with nucleon quantum numbers. Those 
 with positive parity and spin $1/2$ 
can also be represented in terms of scalar and 
pseudoscalar diquarks
\be
\label{ioffe_ps}
\eta_{1,2} = (2,4) \left\{ \epsilon_{abc} (u^a C d^b)\gamma_5 u^c 
 \mp \epsilon_{abc} (u^a C\gamma_5 d^b) u^c \right\}.
\ee
Nucleon correlation functions are defined by $\Pi^N_{\alpha\beta}(x)=\langle 
\eta_\alpha(0)\bar\eta_\beta(x) \rangle$, where $\alpha,\beta$ are the Dirac 
indices of the nucleon currents. In total, there are six different nucleon 
correlators: the diagonal $\eta_1\bar\eta_1,\,\eta_2\bar\eta_2$ and 
off-diagonal $\eta_1\bar\eta_2$ correlators, each contracted with either the 
identity or $\gamma\cdot x$. Let us focus on the first two of these correlation
functions (for more detail, see\cite{SS_98} and references therein).

\begin{figure}[htb]
\includegraphics[width=8cm]{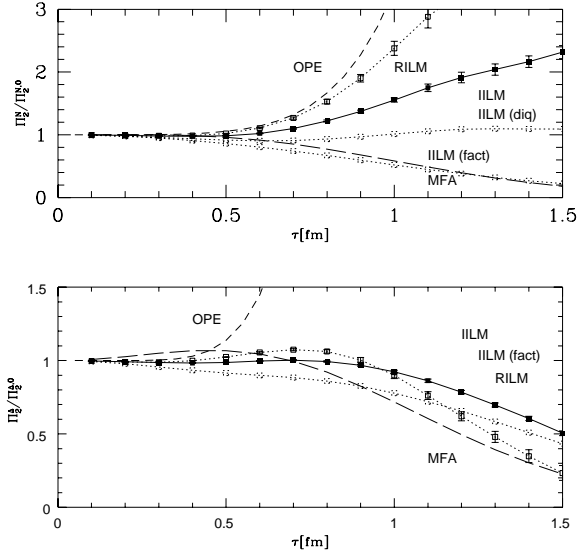}
\vspace{-3mm}
\caption{\label{fig_bar_cor}
Nucleon and delta correlation functions $\Pi_2^N$ and $\Pi_2^\Delta$. Curves
labeled as in Figs.on mesonic correlators.}
\end{figure}

The correlation function $\Pi_2^N$ in the interacting ensemble is shown in 
Fig.~\ref{fig_bar_cor}. 
 The fact that 
the nucleon in IILM is actually bound
 can also be demonstrated by comparing the full nucleon 
correlation function with that of three non-interacting quarks (the cube of 
the average propagator). The full correlator is 
significantly larger than the non-interacting one.

There is a significant enhancement over the perturbative
contribution which is nicely described in terms of the nucleon contribution. 
Numerically, we find\footnote{Note that this value corresponds to a relatively 
large current quark mass $m=30$ MeV.} $m_N=1.019$ GeV. In the random ensemble, 
we have measured the nucleon mass at smaller quark masses and found $m_N=0.96
\pm 0.03$ GeV. The nucleon mass is fairly insensitive to the instanton ensemble.
However, the strength of the correlation function depends on the instanton 
ensemble. This is reflected by the value of the nucleon coupling constant, 
which is smaller in the IILM.
In\cite{SSV_94} we studied all six nucleon correlation functions. We showed 
that all correlation functions can be described with the same nucleon mass and 
coupling constants.

The fitted value of the threshold is $E_0\simeq 1.8$ GeV, indicating that there
is little strength in the ``three quark continuum'' (dual to higher
resonances
in the nucleon channel).
 A significant part of this interaction  was traced down to
the strongly attractive {\em scalar diquark} channel.
 The nucleon (at least in 
IILM) is a strongly bound diquark, plus a  loosely bound  third quark. 
The properties of this diquark picture of the nucleon continue to be
disputed
by phenomenologists. We will return to diquarks in the next section,
where
they will become Cooper pairs of Color Super-conductors.

In the case of the $\Delta$ resonance, there exists only one independent Ioffe
current, given (for the $\Delta^{++}$) by
\be
\eta^\Delta_\mu =  \epsilon_{abc} (u^a C\gamma_\mu u^b)  u^c.  
\ee
However, the spin structure of the correlator $\Pi^\Delta_{\mu\nu;\alpha\beta}
(x)=\langle\eta^\Delta_{\mu\alpha}(0) \bar \eta^\Delta_{\nu\beta}(x)\rangle$
is much richer. In general, there are ten independent tensor structures, but 
the Rarita-Schwinger constraint $\gamma^\mu \eta_\mu^\Delta=0$ reduces this 
number to four. 

The mass of the delta resonance is too large in the random model, but closer to
experiment in the unquenched ensemble. Note that,similar to the nucleon, part 
of this discrepancy is due to the value of the current mass. Nevertheless, the 
delta-nucleon mass splitting in the unquenched ensemble is $m_\Delta-m_N=409$ 
MeV,  larger but comparable to the experimental value 297 MeV.
It mostly comes from the {\em absent scalar diquarks} in $\Delta$ channel.

\section{The Phases of QCD}
\subsection{The Phase Diagram}

In this section we discuss
QCD in extreme conditions, such as
 finite temperature/density.  Let me first 
emphasize why it is interesting and instructive to do. It is
  not simply to practice
once again the semi-classical or perturbative methods
 similar to what have been done before in vacuum. What we are looking for
 here
are {\em new phases} of QCD (and related theories), namely 
new self-consistent solutions which differ qualitatively from
what we have in the QCD vacuum. 

One such phase occurs at high enough temperature $T>T_c$: it is known
as
Quark Gluon Plasma (QGP). It is a phase  understandable in
terms of basic quark and gluon-like excitations\cite{Shu_80}, without
confinement
and with unbroken chiral symmetry in the massless limit\footnote{
It does not mean though, that it is a simple issue to understand even
the high-T limit of QCD, related to non-perturbative 3d dynamics.}.
 One of the main goals of heavy ion
program, especially at new the dedicated Brookhaven facility RHIC, is to
study transitions to this phase.

 Another one, which has been getting much attention recently, is the direction
of finite density. Very robust Color Superconductivity was found to
be the case
here. Let me also mention
one more frontier 
 which has not yet attracted sufficient
attention: namely a transition (or many transitions?) as the number of
light flavors $N_f$ grows. The minimal scenario includes a transition
from the usual hadronic phase to a more unusual QCD phase,
the $conformal$ one, in which there are no 
particle-like excitations and correlators
   are power-like in the infrared. Even the position of the critical
   point is unknown. 
The main driving force of these studies is the
intellectual challenge it provides.

\begin{figure}[ht]
\vskip .05in
 \centering
 \includegraphics[width=2.2in, angle=270]{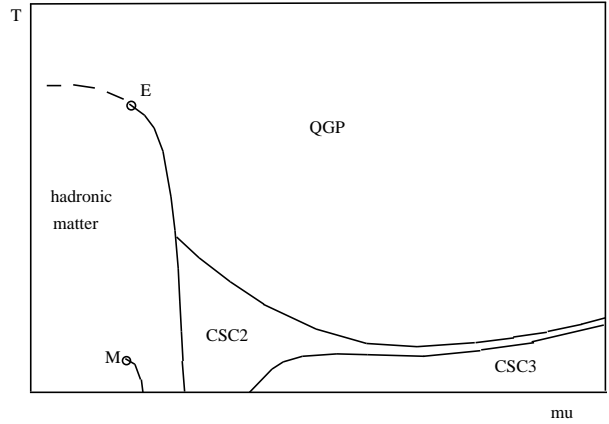}
\vskip 0.1in
\caption{
\label{fig_phases_th}
 Schematic phase diagram of QCD, in temperature T- baryon chemical
potential $\mu$ plane. E and M show critical endpoints of first
order transitions: M (from multi-fragmentation) is that for liquid-gas
transition in
nuclear matter. The color superconducting phases, CSC2 and   CSC3 are
explained in the text.
 }
\end{figure}

  The QCD phase diagram as we understand it now
is  shown in Fig
\ref{fig_phases_th}(a), in the baryonic chemical potential $\mu$
(normalized per quark, not per baryon) and the temperature T
plane. Some part of it is old: it has the hadronic phase at small values
of both parameters, and QGP phase at large T,$\mu$.

The phase transition line separating them most probably does not really
start at $T=T_c,\mu=0$ but at an ``endpoint'' E, 
a remnant
  of the so called QCD tricritical point which QCD has in the chiral
(all quarks are massless) limit. 
Although we do not know where it is\footnote{Its position is very sensitive to
the precise
value of the strange quark mass $m_s$}, we hope to find it one day in experiment. The proposed ideas rotate around the fact that the
 order parameter, the VEV of the sigma meson, is
at this point truly massless, and creates a kind of ``critical opalecence''. 
Similar phenomena were predicted and then indeed observed at the
endpoint of another line (called M from multi-fragmentation), separating
liquid nuclear matter from the nuclear gas phase. 

The 
large-density (and low-T) region
looks rather different from what was shown at
 conferences just a year ago:  two new 
 Color Super-conducting phases appear there.  Unfortunately
  heavy ion collisions do not cross this part
of the phase diagrams and so  it belongs to  neutron star
physics.

   Above I mentioned an  approach to high density starting from the vacuum.
One can also work out in the opposite direction, starting from 
 very large densities and going down. Since the
electric part of one-gluon exchange is screened,
and therefore the Cooper pairs appear due to magnetic forces.
It  is
 interesting by itself, as a rare example: 
one has  to take care of {\em time delay effects} of the interaction.
 The result is  indefinitely growing gaps
  at  large $\mu> 10 GeV$, as \cite{Son}
$ \Delta \sim \mu exp( -{3\pi^2 \over \sqrt{2}g(\mu)})$.

\subsection{Finite Temperature transition and Large Number of Flavors}

  There is no place here to discuss  in detail the
 rather extensive lattice data available now, and I only 
 mention some results related to
 instantons. In the vacuum a quasi-random
set of instantons leads to chiral symmetry breaking and quasi-zero
modes: but what in the same terms does
 the high-T phase look like?

The simplest solution would be just  $suppression$  of instantons
at $T>T_c$, and at some early time people thought this is what actually
happens. However, it should not be like this because the Debye
screening which is killing them only appears at $T=T_c$. Lattice
data works have also found no depletion of the instanton density
up to  $T=T_c$.

\begin{figure}[h]
\vskip .05in
 \centering
 \includegraphics[width=3.in]{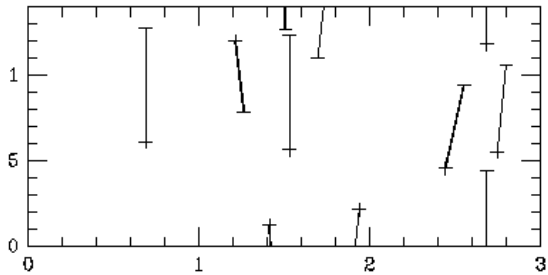}
  \vspace{-.05in}
  \caption{\label{molec}
Typical configuration from instanton liquid simulation, at $T>T_c$.
Lines indicate the direction in which quark propagators are the largest.
Clear pairing of instantons and instantons are observed: the pairs
tend to have the same spatial position and being separated mostly by
Euclidean time.
  }
\end{figure}

  On the other hand,
the absence of the condensate and quasi-zero modes
implies that the ``liquid" is now broken into finite pieces.
The simplest of them are pairs, or the instanton-anti-instanton
molecules. This is precisely what instanton simulations have
found\cite{SS_98},
see fig.\ref{molec}.
Whether it is indeed so on the lattice is not yet clear: 
nice molecules were located, but the evidence 
for the molecular mechanism 
of chiral restoration is still far from being  convincing.
(No alternative I am aware of have been so far  proposed, though.)

  The results of IILM simulations with
variable number of flavors $N_f=2,3,5$\footnote{The case 
$N_f=4$ is omitted because in this case it is very hard to determine 
whether the phase transition happens at $T>0$.} flavors with equal 
masses can be summarized as follows. 
 For $N_f=2$ there is a second order phase transition 
which turns into a line of first order transitions in the $m-T$ plane
for $N_f>2$. If the system is in the chirally restored phase ($T>T_c$) 
at $m=0$, we find a discontinuity in the chiral order parameter if 
the mass is increased beyond some critical value. Qualitatively, the 
reason for this behavior is clear. While increasing the temperature 
increases the role of correlations caused by fermion determinant, 
increasing the quark mass has the opposite effect. We also observe 
that increasing the number of flavors lowers the transition temperature. 
Again, increasing the number of flavors means that the determinant
is raised to a higher power, so fermion induced correlations become
stronger. For $N_f=5$ we find that the transition temperature drops
to zero and the instanton liquid has a chirally symmetric ground state, 
provided the dynamical quark mass is less than some critical value. 
  Studying the instanton ensemble in more detail shows that in this 
case, all instantons are bound into molecules.

   Unfortunately, little is known about QCD with large numbers of
flavors from lattice simulations. There are data
by the Columbia group 
for $N_f=4$. The most important result is that chiral symmetry
breaking effects were found to be drastically smaller as compared 
to $N_f=0,2$. In particular, the mass splittings between chiral
partners such as $\pi-\sigma,\,\rho-a_1,\,N(\frac{1}{2}^+)-N(\frac{1}{2}^-)$, 
extrapolated to $m=0$ were found to be 4-5 times smaller. This 
agrees well with what was found in the interacting instanton model:
more work in this direction is certainly needed.

\subsection{High Density and Color Superconductivity}

This topic is covered in detail by Prof.Alford at this school, so
I only add few remarks to his lectures.

Although the idea of color superconductivity originates from 70's,
the field of high density QCD was in the dormant state for long
time till two papers \cite{RSSV,ARW} (posted on the same day) in 1998
have claimed gaps about 100 times larger than previously thought.
The field is booming since, as one can see from about 250 citations in
2 years those papers got.

 Then-Princeton group (Alford-Rajagopal-Wilczek) have been thinking
about
different pairings from theory perspective, but our (Stony Brook) team
(Rapp,Schafer,ES,Velkovsky) had started from the impressive qq pairing
phenomenon found theoretically  
\cite{SSV_94}  in the instanton liquid model
  {\em inside the nucleon}. As explained above, we have found it to
be,
roughly speaking, a small drop of CS matter, made of
one Cooper pair of sort
(the  {\em ud scalar diquark}) and one massive quark\footnote{As opposed to $\Delta$
  (decuplet) baryons, which is a small drop of ``normal'' quark
matter, without scalar diquarks.} .
T.Schafer heroically attempted numerical simulations of the instanton liquid
model
at finite $\mu$: although he was not very successful\footnote{for the
same reason as lattice people cannot do it: the fermionic
determinant is not real.} he found out strange ``polymers'' made of
instantons connecting by 2 through going quark lines. It take us some
time
to realize we see paths of condensed diquarks! It was like finding
superconductivity by watching electrons moving on your computer screen.

The main point I would like to emphasize here is that the $qq$ pairing
 of such diquarks have in fact  deep dynamical roots: it follows
 from the same basic dynamics
as the  ``superconductivity'' of the QCD vacuum, the chiral
 ($\chi$-)symmetry breaking. These
spin-isospin-zero
   diquarks are related to pions, as we will see below.

The most straightforward argument for deeply bound diquarks came from the 
bi-color ($N_c=2$) theory: in it the scalar diquark is degenerate
with pions.  By
 continuity from $N_c=2$ to $3$, 
 a trace of it
 should exist in real QCD\footnote{
Instanton-induced interaction strength in diquark channel is
$1/(N_c-1)$ of that for $\bar q\gamma_5 q$ one. It is the same at
$N_c=2$, zero for large  $N_c$, and is   exactly in
 between for $N_c=3$.}.

Instantons create  the following amusing {\em  triality}: 
there are three attractive channels 
 which compete: (i)
the
{instanton-induced} attraction in
$\bar q q $ channel
leading to  $\chi$-symmetry breaking.
(ii) The
{instanton-induced} attraction in
$ q q $ which leads to color superconductivity.
(iii)
The
{\em light-quark-induced} attraction of $\bar I I $, which
leads to pairing of instantons
into {``molecules''} and a Quark-Gluon Plasma (QGP) phase without 
$any$ condensates.

At very high density we also can find {\em arbitrarily dilute
instanton liquid}, as shown recently in \cite{SSZ}. The reason
it cannot exist in vacuum or high T is that if instanton density goes
below
some critical value, the cannot be any condensate. (The system then
breaks into instanton molecules or other clusters and chiral symmetry 
is restored.) However at high density the superconducting condensate
can be created perturbatively as well (we mentioned it above) and
there is no problem. The dilute instantons interact by exchanging very 
light $\eta'$ (which would be massless without instantons): one can
calculate effective Lagrangian, theta angle dependence etc.  

{\bf Bi-color QCD: a very special theory  }
One reason it is special (well known to to the lattice community): its
fermionic determinant is $real$ even for non-zero $\mu$,
 which makes simulations possible.
 However the major interest in this theory
is related the so called {\em Pauli-Gursey symmetry}.
We have argued above that pions and diquarks appear at the same
one-instanton
level, and are so to say brothers.
In bi-color QCD they becomes identical twins:  due to the additional
symmetry mentioned the diquarks are
{\em degenerate} with mesons.

In particular,  chiral symmetry breaking is done like this
$SU(2N_f)\rightarrow Sp(2N_f)$, and
 for $N_f=2$ the coset 
$ K=SU(4)/Sp(4)=SO(6)/SO(5) =S^5$.
 Those 5 massless modes are
 pions plus the scalar diquark $S$ and its anti-particle $\bar S$. 

Vector diquarks are degenerate with vector mesons, etc. Therefore,
the scalar-vector splitting is in this case
about twice the constituent quark mass, or about 800 MeV. It should be
compared to binding in the ``real'' $N_c=3$ QCD of only 200-300 MeV,
and to zero binding in the large-$N_c$ limit.

The corresponding
sigma model describing this  $\chi$-symmetry breaking
was worked out in\cite{RSSV}: for further development 
see\cite{KST}.  As argued in \cite{RSSV}, in this theory
the
critical value of the transition to Color Superconductivity is simply
$\mu=m_\pi/2$, or zero in the chiral limit. The  diquark condensate is 
  just a rotated $<\bar q q>$ one, and the gap is the constituent quark mass.
Recent lattice
works  \cite{2col}  display it in great detail, building
confidence for other cases.

{\bf New studies reveal possible new crystalline phases.}
These phases still have somewhat debatable status, so I have not indicated
them on the phase diagram.

Once again, there were two papers submitted by chance on the same day.
 The ``Stony Brook'' team\cite{RSZ} have found that a ``chiral crystal''
with oscillating $<\bar q q(x)>$
(similar to Overhouser spin waves in solid state) can compete with the BCS
2-flavor superconductor at its onset, or $\mu\approx 400\, MeV$.
The proper position of this phase is somewhere in between the hadronic
phase (with constant $<\bar q q>$) and color superconductor.

 The ``MIT
group''\cite{ABR} have looked at the oscillating superconducting condensate 
 $<q q(x)>$, following earlier works on the so called LOFF
phase in usual superconductors. They
 have found that it is appearing when the difference between Fermi
momenta
of different quark flavors become comparable to the gap.
The natural place for it on the phase
diagram is close to the line at which color superconductivity
disappears because the gap goes to zero.

\section{High Energy Collisions of hadrons and Heavy Ions}

\subsection{The Little Bang: AGS, SPS and now the {RHIC} era}

Let me start with brief comparison of these two magnificent explosions: the Big
Bang versus the Little Bang, as we call heavy ion collisions.

The expansion law is roughly the Hubble law in both,
$v(r)\sim r$ although strongly anisotropic
in the Little Bang.
The Hubble constant tells us the expansion rate today: similarly
radial flow tells us the final magnitude of the transverse velocity.
The acceleration history is not really well measured.
For Big Bang people use distance supernovae, we use $\Omega^-$ which
does not participate at the late stages to learn {\em what was the
velocity earlier}. Both show small dipole (quadrupole or elliptic for
Little Bang) components which has some physics, and who knows maybe we
will see higher harmonics fluctuations later on, like in Universe.
 As we will
discuss below, in both cases the major puzzle is how this large
entropy
has been actually produced, and why it happened so early.

The major lessons we learned from AGS experiments
($E_{LAB}=2-12 A GeV$) are:\\
(i) Strangeness enhancement over simple multiple NN collisions appear
from very low energies, and heavy ion collisions quickly approach
nearly ideal chemical
equilibrium of strangeness. \\
(ii) ``Flows'' of different species, in their radial,directed and elliptical
form, are in this energy domain
driven by collective potentials and absorptions:  they 
are not really flows in hydro sense. All of them strongly diminish by
the high end of the AGS region, demonstrating the onset of
``softness''
of the EoS. Probably it is some precursor of the QCD phase transition.

Several  important lessons came so far from CERN SPS data:\\
(i) Much more particle ratios have been measured there:
overall those show surprisingly good degree of
chemical
equilibration: the chemical freeze-out parameters are tantalizingly
close to the QGP phase boundary.\\
(ii) Dileptons show that radiation spectral density is very different
in dense matter compared to ideal hadronic gas. The most intriguing
data are CERES finding of ``melting of the $\rho$'',
which seem to be transformed into a wide  continuum reaching down
to invariant masses as low as 400 MeV. It puts in doubt ``resonance
gas'' view of hadronic matter at these conditions. Intermediate mass
dileptons
studied by NA50 can be well described by thermal radiation with QGP rates.
 \\
(iii)The impact parameter of $J/\psi$ and $\psi'$ suppression in PbPb collisions
studied by NA50 collaboration shows rather non-trivial behavior. More 
studies are needed, including especially measurements of the open charm
yields, to understand the origin and magnitude of the suppression.

However, during last several months those discussions
have been overshadowed by a list of news from RHIC,
 Relativistic 
Heavy Ion Collider  at Brookhaven National Laboratory. It had
 its first run in summer 2000 and reported recently at  Quark 
Matter 2001 conference 
 \cite{QM01}: many details are discussed in Prof.M.Gylassy's lectures.

A brief summary is as follows. These results have shown that 
 heavy ions collisions (AA) at these energies  
significantly differ  
$both$ from the pp collisions at high energies 
 and  the AA collisions at lower (SPS/AGS) energies. 
The main features of these data are quite consistent with 
 the Quark-Gluon Plasma (QGP) (or Little Bang) scenario, in which entropy is produced promptly and 
subsequent expansion is close to adiabatic expansion of equilibrated 
hot medium. 

(Let me mention here two other pictures of the heavy ion production,
discuss prior to appearance of these data. One is the {\em string picture},
used in event generators like RQMD and UrQMD: they predicted
effectively very soft EoS and elliptic flow decreasing with
energy. The other one is {\em pure minijet scenario}, in which most
secondaries
would come from independently fragmenting minijets. If so, there are
basically no collective phenomena whatsoever.)

Already the very first multiplicity measurements 
 reported by PHOBOS collaboration  \cite{PHOBOS} 
 have shown that particle production per participant 
  nucleon is no longer constant,  
as was the case at lower (SPS/AGS) energies. 
This new component may  be due to long-anticipated {\em pQCD} 
processes, leading to perturbative production of new partons. Unlike 
high $p_t$ processes resulting in visible jets, those must 
be undetectable  {\em ``mini-jets''} 
 with momenta $\sim 1-2\, GeV$. Production and decay 
of such {\em mini-jets} was discussed 
 in Refs \cite{minijets}, also this scenario is the 
basis of widely used event generator HIJING  \cite{HIJING}. 
Its crucial parameter is the {\em cutoff scale} $p_{min}$: if 
fitted from pp data to be 1.5-2 $GeV$, it leads to predicted mini-jet
multiplicity
$dN_g/dy\sim 200$ for central AuAu collisions at $\sqrt(s)=130 \, 
AGeV$. If those fragment independently into hadrons, and are
supplemented 
  by ``soft'' string-decay component, the predicted total multiplicity 
was found to be in good agreement with the 
first RHIC multiplicity data.  Because  partons 
interact perturbatively,  
with their scattering  and radiation being strongly peaked at small
angles, 
  their equilibration is expected to be  relatively long 
  \cite{equilibr}. 
However, new set of  RHIC data reported in  \cite{QM01}  
have provided  serious arguments {\em against} the mini-jet   
scenario, and point toward quite rapid entropy production rate and 
early QGP formation.

(i) If most of mini-jets fragment 
independently, there is 
no {\em collective phenomena} 
such as transverse flow  
related with the QGP pressure. However, it was found  
that those effects are very strong at RHIC. Furthermore, 
STAR collaboration have observed very robust 
{\em elliptic flow}  
 \cite{STAR}, which is in perfect agreement with 
predictions of hydrodynamical model 
 \cite{h2hflow,Kolb} 
 assuming equilibrated QGP with its 
full pressure $p\approx \epsilon/3$ above the QCD phase 
transition. This agreement persists to rather peripheral 
collisions, in which the overlap almond-shaped region of two nuclei 
is only a couple fm thick. STAR and PHENIX data on spectra of 
identified 
particles, especially $p,\bar p$, indicate spectacular radial
expansion, 
also 
in agreement with hydro calculations   
 \cite{h2hflow,Kolb}. 
 (ii) Spectra of hadrons at large $p_t$, especially the $\pi^0$ 
spectra 
 agree well with HIJING for peripheral collisions, but  show  
much smaller yields 
for central ones, with rather 
different, 
(exponential-shaped) spectra. It means  long-anticipated 
{\em  ``jet quenching''}  at large $p_t$ is seen for the first time,
with  
 a surprisingly large suppression factor $\sim 1/5$. 
 Keeping in mind that jets originating 
from the surface outward cannot be quenched, the effect seem to be as 
large 
 as it can possibly be. For that to happen, the outgoing high-$p_t$ 
 jets  should propagate through matter with parton population 
   larger than the abovementioned 
minijet density predicted by HIJING. 
 
(iii) Curious interplay between collective and jet effects have 
also been studied by STAR collaboration, in form  
of elliptic asymmetry parameter $v_2(p_t)$. At large  
transverse momenta $p_t>2\, GeV$ the data 
depart from hydro predictions and levels off. When compared to 
predictions of jet quenching models worked out in \cite{GVW}, 
 they also indicate gluon 
 multiplicity several times larger than HIJING prediction, and are
even consistent with 
its maximal possible value  evaluated from  the final  
  entropy at freeze-out, $(dN/dy)_\pi\sim 1000$.

\subsection{ Collective flows and EoS}

If we indeed have produced excited matter (rather than just a bunch of
partons which fly away and fragment independently), we expect to see
certain collective phenomena. Ideally, those should be quantitatively 
reproduced by relativistic hydrodynamics which is basically just
local
energy-momentum conservation plus the EoS we know from the lattice and 
models. 

The  role  of the QCD phase transition in matter expansion
is  significant. QCD lattice simulations \cite{lattice}
show approximately 1st order
transition. Over a wide range of energy densities $e=.5-1.4 \, GeV/fm^3$ 
the temperature T and pressure p are nearly constant. So
the ratio of pressure to energy density, $p/e$, decreases till 
 a minimum at particular energy density
$e_{sp}\approx 1.4 \, GeV/fm^{3}$, known as the {\em softest point} \cite{HS-freeze}.
Near $e_{sp}$ small pressure gradient  can not
effectively  accelerate the matter and the evolution
stagnates.
However when the initial energy density is well above the QCD phase transition
region,  $p/e\approx 1/3$,  and this pressure drives the collective motion.
The energy densities reached at time $\sim 1 fm/c$ at SPS($\sqrt
s_{NN}=17 \, GeV$) and RHIC ($\sqrt s_{NN}=130 \, GeV$) are about 4 and 8
$GeV/fm^3$,
respectively. 
 We found that at RHIC conditions we are in the latter regime, and
matter accelerates  to $v\sim .2c$ 
$before$ entering the soft domain. Therefore  by freeze-out this motion
changes the spatial
distribution of matter dramatically: e.g. as shown in
  \cite{TS_nut} the initial almond-shape distribution 10 fm/c later looks
like two separated shells, with a little ``nut'' in between.

The simplest way to see hydro expansion is in spectra of particles:
on top of chaotic thermal distributions $\sim
exp(-m_t/T),m_t^2=p_t^2+m^2$
one expect to see  additional broadening due to hydro outward
motion. This effect is especially
large if particles are heavy, since flow with velocity v add momentum
$m v$. 

\begin{figure}[h]
   \includegraphics[width=2.5in]{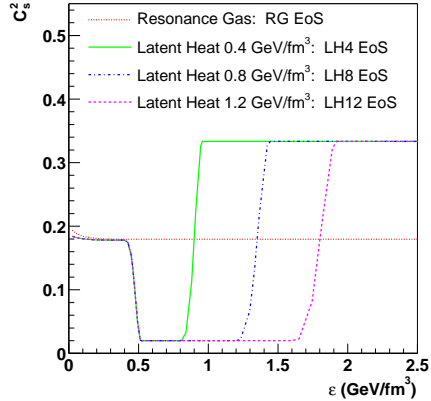}
   \caption[]{
      \label{psCs2}   The EoSs in form squared speed of sound
$c^2=dp/d\epsilon$  
 with variable Latent Heats $.4 GeV/fm^{3}$, $.8GeV/fm^{3}$,..
 labeled as LH4, LH8,..versus the energy density. 
   }
\end{figure}

Derek Teaney \cite{h2hflow} have developed a comprehensive  Hydro-to-Hadrons (H2H) model combines the
hydrodynamical description of the initial QGP/mixed phase ($e>.5 GeV/fm^3$)
 stages,  where
hadrons are not  appropriate degrees of freedom,
 with a hadronic cascade RQMD for
the hadronic stage.
In this 
way, we can include different EoS displaying properties of
the phase transition,
and also incorporate complicated final state interaction at freeze-out.
The  set of EoS  used is shown in Fig.\ref{psCs2}.  

\begin{figure}[h]
\begin{center}
   \includegraphics[width=2.9in]{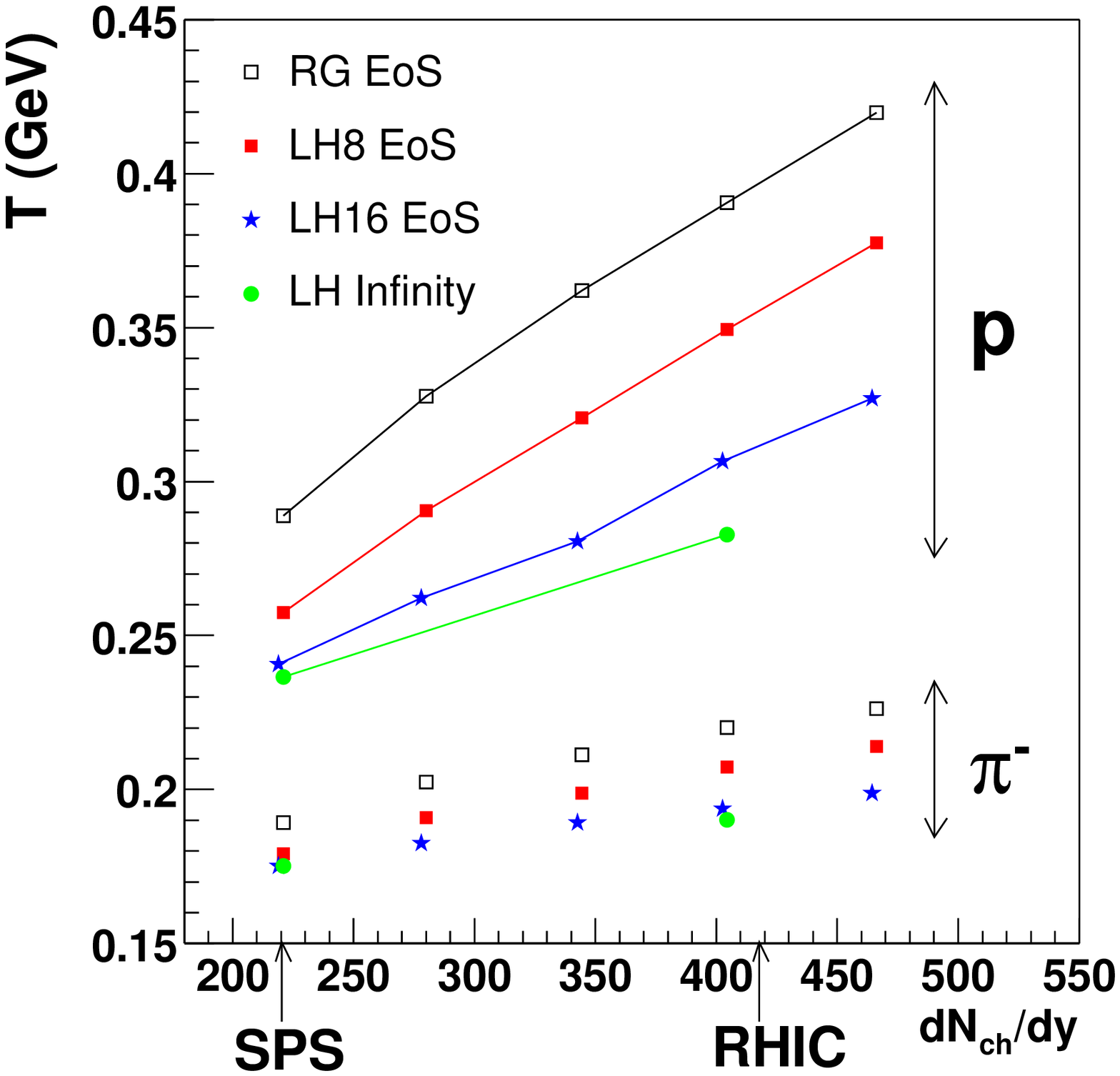}
   \includegraphics[width=2.9in]{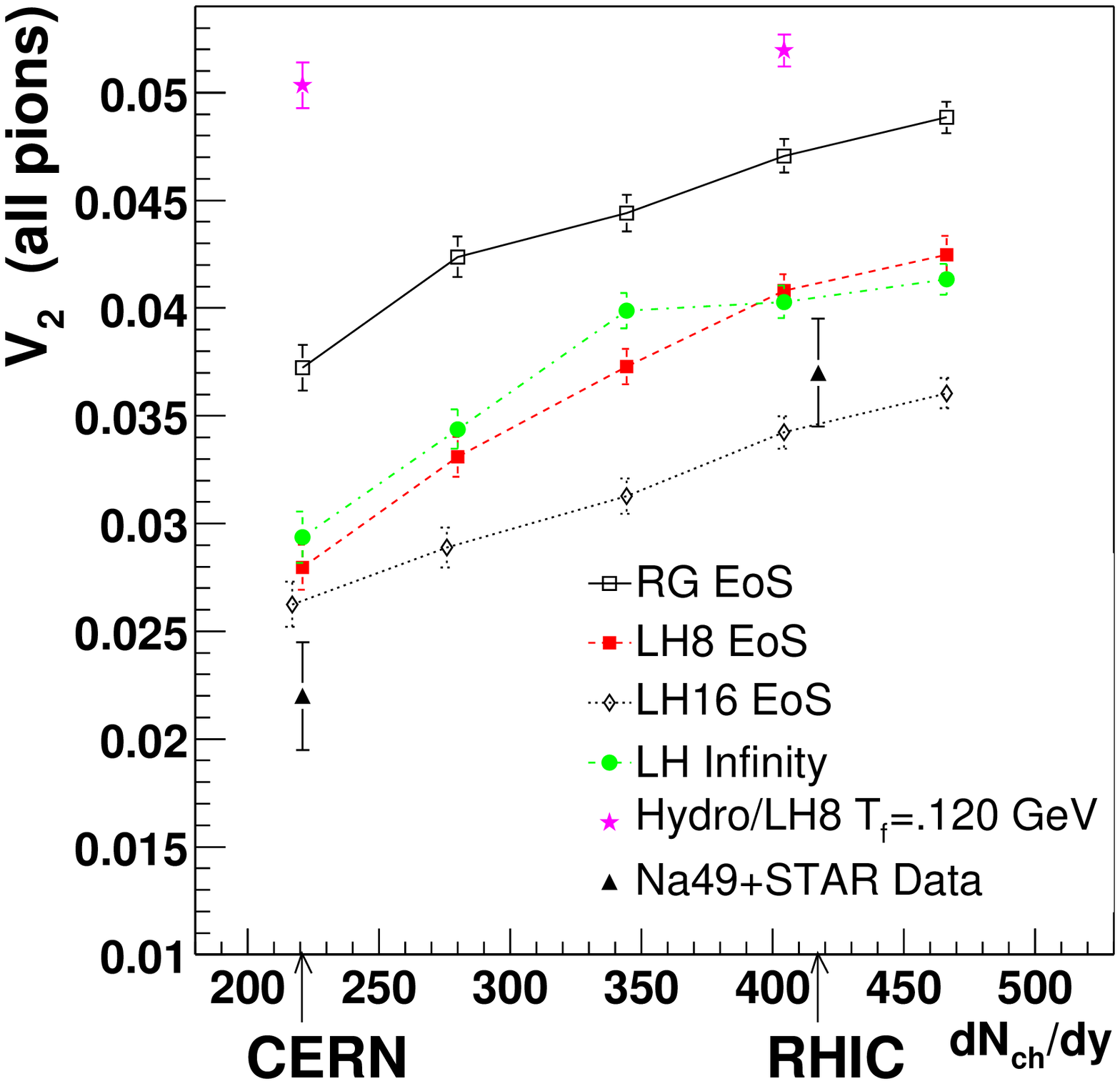}
\end{center}
   \caption[]{
      \label{psPionV2dNdy}
The transverse mass slope  T (a) and elliptic flow  
parameter $V_2$ (b) versus midrapidity (y=0)  charged purticle
multiplicity,
for AuAu collisions with b=6.
   }
\end{figure}

{\em Radial flow} is usually characterized by the slope parameter T:
each particle spectra are fitted to  the form 
$dN/dp_t^2dy\sim exp(-m_t/T), m_t^2=p_t^2+m^2$.
Although we denoted the slope by
 T, it is $not$ the temperature: it incorporates random thermal motion
and collective transverse velocity. The SPS NA49 slope parameters for
pion and protons are 
 shown in Fig. \ref{psPionV2dNdy}(a).
Parameter
T grows with particle multiplicity
 due to increased velocity of the radial flow.
Furthermore, the rate of growth depends on the EoS: the softer it is,
the less growth. The SPS NA49 data correspond to two data points (our
fits
to spectra)  favor the (relatively stiff) LH8 EoS.  
(Details of the fit, discussion of the b-dependence etc see in
\cite{h2hflow}.) It is very important to get these parameters for RHIC, 
especially for heavy secondaries like nucleons and hyperons.

 For $non-central$ collisions the  
overlap region in the transverse plane has an elliptic, ``almond'',
shape, and larger pressure gradient force
 matter to expands preferentially in the direction of the 
impact parameter \cite{Ollitrault}. 
Compared to radial flow, the elliptic flow
is formed earlier, and therefore it measures the early pressure. 
 The {\em elliptic flow} is quantified experimentally
by measuring the azimuthal distributions of the produced particles
and calculating the elliptic flow
parameter $V_{2}=\langle cos(2\phi) \rangle$
where $\phi$ angle is measured with respect to the impact parameter 
direction, around
the beam axis. It appears due to 
 the elliptic $spatial$ deformation of the overlap region in
the nucleus-nucleus collision,  quantified 
by its eccentricity $\epsilon_2 =  <y^{2} - x^{2}> / < x^{2} + y^{2}> $, usually calculated in
Glauber model. Since the effect ( $v_2$) is proportional to
the cause ($\epsilon_2$), the ratio   
 $v_{2}/\epsilon_2$ does not have strong dependence on
the impact parameters b, and this ratio is often used for comparison.
(We would not do that below, in the detailed comparison
to data, because $\epsilon_2(b)$
is not directly measured.

In figure \ref{psPionV2dNdy}(b)  the elliptic flow of the
system is plotted as a function of charged particle multiplicity at an 
impact parameter of 6 fm. Before discussing the energy dependence, 
 let us quantify the
magnitude of elliptic flow at the SPS.
Ideal relativistic hydrodynamics used in earlier works
\cite{Ollitrault,Kolb} generally over-predicts
elliptic flow by about factor 2. Example of such kind is indicated
by a star in figure \ref{psPionV2dNdy}(b): it is our hydro result (with LH8 EoS)
 which has been followed hydrodinamically
till very late stages, the freeze-out temperature $T_f=120 \, MeV$.
By switching to hadronic cascade at late stages, we have more appropriate
treatment of resonance decays and
re-scattering rate, and so one can see that it significantly 
reduces $V_2$, to the range much closer to the data points.

One might thing that one can also do that by simply taking {\em
softer}
EoS, e.g. increasing the latent heat. However, it only happens till
LH16 and then $v_2$ start even slightly increase again. The
explanation
of this non-monotonous behavior is the interplay of the initial ``QGP 
push'' for stiffer EoS, with longer time for hadronic stage available for
softer EoS. We cannot show here details, but it turns out that a given
(experimental) $V_2$ value can correspond to {\em two different
solutions}, one with earlier push and another with the later expansion 
dominating. Coincidentally, STAR data point happen to be right at the
onset of such a bifurcation, close to LH16. 
  The {\em multiplicity dependence} of $V_{2}$ appears simple from 
figure \ref{pspionV2dNdy}(b):
all curve show growth with about the same rate. Note however that such 
growth of $V_2$ from SPS
to RHIC (first predicted in
 \cite{Shu_QM99} where our first preliminary results has been shown) 
Is not shared by most other models. 
In particular, {\em string-based}
models like
 UrQMD 
predicts a decrease by a factor of  $\approx 2$
\cite{UrQMD_Bleicher}. It happens 
 because  strings produce no transverse pressure and so the effective
EoS is super-soft  at
high energies. Models based on {\em independent parton scattering and decay}
(such as HIJING) also predict basically vanishing 
(or slightly negative)\cite{HIJING} $V_2$.

\begin{figure}[h]
   \vskip 0.2in
   \includegraphics[width=2.9in, angle=0]{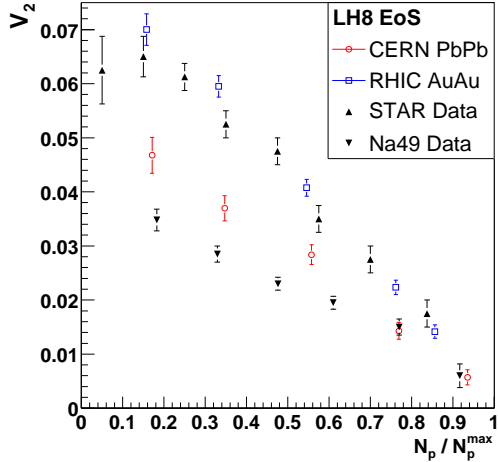}
   \vskip 0.2in
   \caption[]{
      \label{psBscanV2particip}
      v2 versus impact parameter b, described experimentally by the
number of participant nucleons, for RHIC STAR and SPS NA49 experiments. 
 Both are compared to our results, for EoS LH8.
   }
\end{figure}
In Fig.\ref{psBscanV2particip} we show how  our results
compare  with data as a function of impact parameter. One can see
that the agreement becomes much better at RHIC. Furthermore, one may
notice that deviation from linear dependence we predict
becomes visible at SPS for more peripheral
collisions with $N_p/N_p^{max}<0.6$ or so, 
while at RHIC only the most peripheral point, with
$N_p/N_p^{max}=0.05$
show such deviation. This clearly shows that hydrodynamical regime in
general
works much better at RHIC.

In summary, the
 flow phenomena observed at RHIC are stronger than at SPS. It is in
complete agreement with the QGP scenario. All data on elliptic and
radial flow can
be nicely reproduced by the H2H model. Furthermore, we are able to 
 restrict
the
 EoS, to those  with the  latent heat about .8 $GeV/fm^3$.

\subsection{How QGP happened to be produced/equilibrated so early?}

One possible solution to the puzzle outline above
can be a {\em significantly  
 lower cutoff scale} in AA collisions, as compared to
$p_{min}=1.5-2\,GeV$ 
fitted from the pp data. That increases perturbative cross sections, 
both due to smaller momenta transfer and larger coupling constant. 
As I argued over the years, the QGP is a new phase of QCD which is 
 {\em qualitatively different} from the QCD 
vacuum: 
   therefore the  cut-offs of pQCD may  have entirely different values 
and be determined by different phenomena. 
 Furthermore, since QGP is a plasma-like phase which screens itself 
perturbatively \cite{Shu_80}, one may think of a   cut-offs to    
 be determined {\em self-consistently} from resummation of
perturbative 
effects. These ideas known as {\em self-screening} or 
{\em 
initial state saturation} were discussed in Refs. 
 \cite{equilibr}. Although the scale in question grows with  
temperature or density, {\em just above $T_c$} it may actually be   
{\em smaller} than the value 1.5-2 GeV we observe in the vacuum.  
Its first experimental manifestation may be dropping of the so called 
``duality scale'' in the observed 
dilepton spectrum, see discussion in \cite{RW}. 
 
 Another alternative to explain large gluon population at RHIC would
be an existence of 
more rapid multi-gluon production processes.
Let us consider an alternative   $non-perturbative$ scenario 
based entirely on non-perturbative processes involving 
  {\em instantons} and {\em sphalerons} \cite{my_sph}.
But before we do that, we have to take a look at hadronic collisions
and
briefly review few recent papers on the subject.

\begin{figure}[t]  
  \includegraphics[width=6cm]{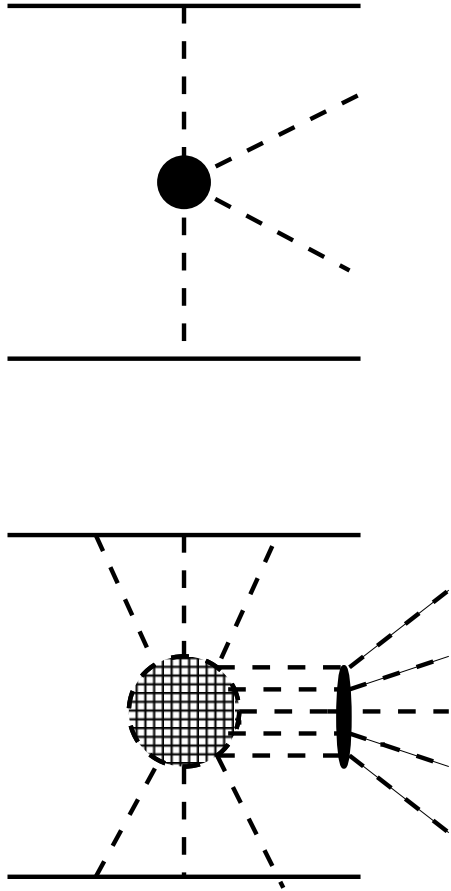} 
   \caption[] 
  { 
   \label{fig} 
(a) A typical inelastic perturbative process: two t-channel gluons
collide, 
producing 
a pair of gluons; (b) Instanton-induced inelastic process incorporate 
collisions 
of multiple t-channel gluons with the  instanton (the shaded circle), 
resulting in multi-gluon production. The intermediate stage of the 
process, 
indicated by 
the horizontal dashed lines, corresponds to a time when outgoing glue 
is 
in the form of coherent field configuration - the $sphaleron$. Since 
this part of the process corresponds to motion above the barrier, it 
does not enter the calculation of the cross section, but is only 
needed 
for prediction of the inclusive spectra, multiplicities etc. 
  } \end{figure}

 At $s>10^3 GeV^2$  hadronic cross sections as $\bar p p,pp,\pi p,
Kp$, 
   $\gamma N$ and even $\gamma\gamma$   
 slowly grow  with the collision 
energy s. This behavior can be well parameterized 
by  {\em soft Pomeron} phenomenology, 
 but we will only use its logarithmically  
growing part   
\be \sigma_{hh'}(s)= \sigma_{hh'}(s_0)+ log (s/s_0) X_{hh'}\Delta
+... 
\ee  
 ignoring both the 
higher powers of log(s) and decreasing Regge terms. 
We will use those two parameters from PDG-2000 recent fits, 
 the intercept  
 and its coefficient in  
$pp,\bar p p$ collisions, 
$ \Delta=  \alpha(0)-1=  0.093(2), \, X_{NN}= 18.951(27)\ mb. $ 
Note a 
 qualitative  difference between constant and logarithmically growing
parts of t\
he 
cross section. The former can be explained by prompt color 
$exchanges$, as suggested by Low and Nussinov  long ago. It 
nicely correlates with flux tube picture of the final state. 
section.) 
The $growing$ part of the cross section 
 cannot be generated by t-channel color exchanges and is 
 associated with  processes promptly  producing some objects, 
with  log(s) coming  from the  longitudinal phase space. 
 In pQCD  it is  {\em gluon} production, by processes like the one 
shown in Fig.\ref{fig}(a). If  
iterated in the t-channel 
in ladder-type fashion, the result is  approximately a BFKL pole 
 \cite{BFKL}. 
Although 
the power  predicted is much larger than $\Delta$ mentioned,  
 it seem to be consistent with much stronger growth seen in hard
processes at 
HERA: thus it is therefore sometimes called 
the ``hard pomeron''.  
 The physical origin of cross section growth remains an outstanding
open 
problem: 
neither the  perturbative resummations 
nor many 
non-perturbative models are really quantitative. 
It is hardly surprising, since  scale at which  soft Pomeron 
operates 
(as seen e.g. from the Pomeron slope $\alpha'(0)\approx 1/(2 \,
GeV)^2$) 
is also  the ``substructure scale'' mentioned above.

Recent application of the instanton-induced dynamics 
to this problem 
have been discussed in several papers  
\cite{pom_inst}. 
 Especially relevant for this Letter are two last works which 
 use  insights obtained a decade ago in 
discussion of instanton-induced processes in electroweak theory 
\cite{weakinst}, and  
 the growing part of the hh cross sections were  ascribed to
multi-gluon 
production via instantons, 
see  Fig.\ref{fig}(b). 
 Among qualitative features of this theory 
is the explanation of why no odderon appears (instantons are SU(2) 
objects, in which quarks and antiquarks are not really distinct), 
an explanation of the small power $\Delta$ (it is proportional to 
``instanton diluteness parameter'' $n\rho^4$ mentioned above),  
the small size of the soft Pomeron 
(governed simply by small size of instantons  
$\rho\sim 1/3 \, fm$).  
Although instanton-induced amplitudes contain small 
``diluteness'' 
factor, there is no extra penalty for production of new  
gluons: thus one should  expect instanton effects to beat perturbative 

amplitudes 
of sufficiently high order. This generic idea is also behind the 
present 
work, dealing with prompt multi-gluon production.

  Technical description of the process can be split into two stages.  
The first (at which one evaluates the probability)  
is the motion {\em under the barrier}, and it is  described by  
Euclidean paths approximated by   
instantons. Their 
interaction with the high energy 
colliding partons results in some energy deposition and subsequent  
 motion  {\em over the barrier}. At this second stage 
the action is real, and the factor  $exp(iS)$ 
 does not affect the probability, and we  only need to consider 
it for final state distributions. 
The relevant Minkowski paths start with  
configurations  close to the 
QCD analogs of electroweak {\em sphalerons} \cite{Manton},  
 static spherically symmetric clusters 
 of gluomagnetic field which satisfy the 
 Yang-Mills equations. (Those can be  
obtained from known electroweak solutions 
 in the limit of infinitely large Higgs self-coupling.) Their mass in 
QCD is 
\be M_{sph}\approx {30 \over g^2(\rho) \rho}\sim  2.5 \, GeV \ee 
Since those field configurations 
are close  to  classically unstable saddle point 
at the top of the barrier, they roll downhill and develop gluoelectric 
fields.  When both become  weak enough, 
solution can be decomposed  into perturbative     
gluons. This part of the process can also be studied directly from 
classical 
Yang-Mills equation: for electroweak sphalerons it has been done in 
Refs\cite{sphaleron_decay}, calculation 
for its QCD version is in progress \cite{CS}. 
While rolling,  the configurations tend to forget the initial 
imperfections (such as a non-spherical shapes) since there is 
only one basic instability path downward:  so 
 the resulting fields 
should be  nearly perfect spherical expanding shells. 
 Electroweak sphalerons  decay into approximately 51 
W,Z,H quanta, of which only about 10\%  are Higgses, which carry only 
4\% of energy. Ignoring those, one can  
estimate  mean gluon multiplicity per  sphaleron decay, 
by simple re-scaling of the coupling constants: the result gives 3-4
gluons.
Although this number is not large, it is important to keep in mind that
they appear as a coherent expanding shell of strong gluonic field.   

It has been suggested in \cite{my_sph} that if sphaleron-type object
are
copiously produced, with or instead of $p\sim 1 GeV$ minijets, they 
may significantly increase the entropy produced and speed up the
equilibration process.

{\bf Acknowledgments.} 
It is a pleasure to thank the organizers of the school for their invitation:
it was indeed a very interesting and successful one.
This work was supported in parts by the US-DOE grant 
DE-FG-88ER40388.

\end{document}